\documentclass[apj]{emulateapj}

\usepackage{graphicx,times}
\usepackage[caption=false]{subfig}
\newcommand{\be}{\begin{equation}}
\usepackage{threeparttable}
\usepackage{booktabs}
\newcommand{\ee}{\end{equation}}
\newcommand{\bea}{\begin{eqnarray}}
\newcommand{\eea}{\end{eqnarray}}

\usepackage{enumerate}
\usepackage{amsmath}
\usepackage{cases}
\usepackage{longtable}
\usepackage{hyperref}
\usepackage{amsmath,bm}
\usepackage{amssymb}
\usepackage{natbib}
\usepackage{morefloats}
\usepackage{multirow}
\usepackage{array}
\usepackage{verbatim}

\begin{document}

\title{On the formation of density filaments in the turbulent interstellar medium}

\author{Siyao Xu\altaffilmark{1,2}, 
Suoqing Ji\altaffilmark{3},
and Alex Lazarian\altaffilmark{1} }

\altaffiltext{1}{Department of Astronomy, University of Wisconsin, 475 North Charter Street, Madison, WI 53706, USA; 
sxu93@wisc.edu,
lazarian@astro.wisc.edu}
\altaffiltext{2}{Hubble Fellow}
\altaffiltext{3}{TAPIR \& Walter Burke Institute for Theoretical Physics, California Institute of Technology, Pasadena, CA 91125, USA;
Sherman Fairchild Fellow; suoqing@caltech.edu}

\begin{abstract}

This study is motivated by recent observations on ubiquitous interstellar density filaments and 
guided by modern theories of compressible magnetohydrodynamic (MHD) turbulence. 
The interstellar turbulence shapes the observed density structures. 
As the fundamental dynamics of compressible MHD turbulence, 
perpendicular turbulent mixing of density fluctuations 
entails elongated density structures aligned with the local magnetic field, 
accounting for low-density parallel filaments seen in diffuse atomic and molecular gas. 
The elongation of low-density parallel filaments depends on the turbulence anisotropy. 
When taking into account the partial ionization, we find that the minimum width of 
parallel filaments in the cold neutral medium and molecular clouds is determined by the neutral-ion decoupling scale perpendicular to magnetic field. 
In highly supersonic MHD turbulence in molecular clouds, 
both low-density parallel filaments due to anisotropic turbulent mixing and 
high-density filaments due to shock compression exist.

\end{abstract}


\section{Introduction}

Observations reveal that filamentary density structures widely spread in the interstellar medium (ISM), 
including both diffuse media 
(e.g., \citealt{McC06,Cl14,Pla16,Kal16})
and highly fragmented molecular clouds (MCs)
(e.g., \citealt{SE79,Will00,And10}).
Moreover, low-density filaments in diffuse media and diffuse regions of MCs
preferentially align with the magnetic field, 
whereas dense filaments in MCs tend to be perpendicular to the magnetic field 
\citep{Gol08,Kal16, Pla16,Pln16}.
The former provides the information on the interstellar magnetic field. The latter is important for understanding the star formation process,
as dense filaments in MCs coincide with birthplaces
of protostellar cores
\citep{And14,Kon15,Mar16}.
Besides its significance in interstellar processes, HI filaments parallel to magnetic fields can also be useful for  
studies of cosmic microwave background polarization
\citep{Cl15}.

As the ISM is turbulent
\citep{Armstrong95,CheL10}, 
understanding the turbulence properties 
is essential for explaining the magnetic field and density structures in the ISM. 
Important strides have been made towards advancing the theories of magnetohydrodynamic (MHD) turbulence 
concerning e.g. the dynamics and statistics of MHD turbulence
(\citealt{GS95}, hereafter GS95; \citealt{LV99}, hereafter LV99),
MHD turbulence in a compressible medium
(\citealt{LG01}, henceforth LG01; \citealt{CL02_PRL,CL03})
and in a partially ionized medium 
(LG01; \citealt{LVC04,XuL15,Xuc16}),
and their numerical testing 
\citep{MG01,CLV_incomp,KL09,KL12}.
With a general applicability in essentially all branches of astrophysics involving fluid dynamics 
\citep{Mckee_Ostriker2007},
the theoretical developments of MHD turbulence provide new insights in many long-standing problems 
(e.g., \citealt{YL02,LY14,XL17,XuZ17,XZg17}).
In particular, the turbulent reconnection of magnetic fields inherent in MHD turbulence 
(LV99)
was introduced to resolve the magnetic flux problem in star formation
\citep{Laz05,Sant10, LEC12}, 
and the turbulence anisotropy has been employed to develop a
new velocity gradient technique for measuring interstellar magnetic fields 
\citep{Gon17,Yu17, LY18}.

In the multi-phase ISM, 
the interstellar turbulence induces density fluctuations and influences density structures in different phases. 
In diffuse warm phases, the turbulence is subsonic to transonic
\citep{Hi08, Chep10, Gae11, Bur12},
while the turbulence in dense cold phases is supersonic 
\citep{Zu74,Zuck74,Lars81,HB04}.
In accordance with distinctive turbulence regimes, 
we naturally expect that the resultant density structures differ in different phases.
This has been confirmed by statistical studies of turbulent densities 
\citep{BLC05,KL07,Lad08,Burk09,Burk10, Col12,Fed12, Bur15}. 
In compressible MHD turbulence, 
density fluctuations are passively regulated by Alfv\'{e}nic modes, 
which are responsible for the dynamics of MHD turbulence, 
and thus present the same statistical features as turbulent velocities 
(LG01, \citealt{CL03}). 
But when the compressible MHD turbulence is highly supersonic, 
besides the density fluctuations associated with Alfv\'{e}n modes,
shocks driven by supersonic turbulent flows can produce additional density structures, 
which are characterized by large density contrasts on small scales due to the shock compression. 
The fact of different density structures arising in different turbulence regimes in the multi-phase ISM
has been shown by overwhelming observations
(see e.g., \citealt{Armstrong95,CheL10}; reviews by \citealt{Laz09rev,HF12} and references therein),
and has been applied by
\citet{XuZ16,XuZ17}
to modeling the Galactic electron density fluctuations and interpreting 
interstellar rotation measure fluctuations and 
scattering measurements of pulsars.

In this paper, based on the established and numerically tested theories of compressible MHD turbulence, 
we will investigate the physical origin of density filaments in the multi-phase ISM. 
In Section 2, we introduce the properties of compressible MHD turbulence in the ISM. 
In Section 3, we focus on the formation mechanism of low-density parallel filaments in both 
sub- and super-sonic interstellar turbulence, and present both the analytical theory 
and numerical tests. 
In Section 4, we briefly discuss the formation mechanism of dense filaments in the supersonic turbulence of MCs. 
The further discussion and summary follow in Sections 5 and 6, respectively.

\section{MHD turbulence in the multi-phase ISM}

The ISM is compressible.
Theories of incompressible strong MHD turbulence 
(GS95; LV99)
have been extended for studying compressible MHD turbulence, on the basis that 
(1) the solenoidal Alfv\'{e}nic component of turbulent velocities dominates over the compressible component 
\citep{Bal99,CL02_PRL,KowL10};
(2) Alfv\'{e}nic component remains incompressible and is only marginally coupled with compressible slow and fast modes
(LG01; \citealt{CL02_PRL, CL03});
(3) slow modes passively follow the cascade of Alfv\'{e}n modes and thus have the same anisotropic scaling as Alfv\'{e}n modes, 
while fast modes independently have an isotropic scaling
(LG01; \citealt{CL02_PRL, CL03}). 
Both theoretical arguments 
(LG01)
and numerical evidence by decomposing compressible MHD turbulence into Alfv\'{e}n, slow, and fast modes 
\citep{CL02_PRL, CL03}
justify the validity of the scaling and anisotropy of Alfv\'{e}nic turbulence in a compressible medium.

In the multi-phase ISM with dramatically varying temperatures, 
the compressible MHD turbulence can be sub- or super-sonic in different interstellar phases.

(1) Subsonic to transonic warm ionized medium (WIM) and warm neutral medium (WNM)

The sonic Mach number $\mathcal{M}_\mathrm{s}$ is defined as the ratio of the turbulent speed at the driving scale of turbulence 
to the sound speed. 
$\mathcal{M}_\mathrm{s} \lesssim 1$ measured in the WIM 
\citep{kul87, Haf99, Hi08, Gae11, Bur12}
and in the WNM
\citep{Chep10}
indicates that the turbulence in these diffuse warm phases is subsonic to transonic. 
In compressible MHD turbulence with $\mathcal{M}_\mathrm{s}$ of the order of unity, 
slow modes are passively mixed by Alfv\'{e}nic turbulent velocities
(LG01). 
Therefore, density fluctuations resulting from slow modes are dictated by the dynamics of Alfv\'{e}n modes 
and obey the same scaling as Alfv\'{e}n modes.
The same Kolmogorov spectrum of density fluctuations as that of velocity fluctuations has been confirmed by MHD simulations 
(e.g., \citealt{CL03}).
Observations of the WIM reveal a 
``big power law in the sky" of electron density fluctuations
spanning from $10^{6}$~m  to hundreds of parsecs,  
which is consistent with the Kolmogorov scaling
\citep{Armstrong95,CheL10}.

(2) Supersonic cold neutral medium (CNM) and MCs

The turbulence in colder and denser phases are supersonic with $\mathcal{M}_\mathrm{s} > 1$ 
\citep{Zuc74, Lars81, Chep10}. 
In highly supersonic MHD turbulence, density fluctuations are subjected to not only the Alfv\'{e}nic mixing, 
but also the shock compression. 
The former acts to smooth the density variation, 
whereas the latter produces large density contrasts and results in 
a shallow density spectrum with 
a significant excess of high-density structures at small scales. 
Such shallow density spectra have been confirmed by supersonic MHD simulations 
\citep{BLC05,KL07}, 
and also measured in the CNM and MCs 
\citep{Laz09rev,HF12}.
Importantly, we note that 
even in highly supersonic MHD turbulence, 
the turbulence dynamics is still dominated by Alfv\'{e}n modes.
Different from the density spectrum, 
the spectrum of velocity fluctuations in supersonic MHD turbulence does not significantly deviate from the Kolmogorov scaling
\citep{KowL10}.
Turbulent velocities measured in the 
low-density region of a MC exhibit a Kolmogorov spectrum 
\citep{Qi18}.

The different density structures in the subsonic to transonic turbulence of diffuse warm phases and 
in the supersonic turbulence of dense cold phases hold the key to understanding a variety of phenomena in the ISM, 
e.g., the interstellar scattering of Galactic pulsars
\citep{XuZ17}
and interstellar rotation measure fluctuations 
\citep{XuZ16}, 
as well as the density filaments in the ISM. 
In what follows we first focus on the formation mechanism of low-density filaments 
based on the dynamics of Alfv\'{e}n modes in both subsonic and supersonic interstellar turbulence
(Section \ref{sec:fldpa}). 
Then we briefly discuss the formation of dense filaments in highly supersonic interstellar turbulence (Section \ref{sec: densho}).

\section{Formation of low-density parallel filaments in both subsonic and supersonic interstellar turbulence }
\label{sec:fldpa}

\subsection{Perpendicular turbulent mixing and low-density parallel filaments}
\label{ssec: ptmpf}

In Alfv\'{e}nic turbulence, the turbulent mixing 
of magnetic fields takes place in the direction perpendicular to the {\it local} magnetic field,
which is enabled by the turbulent reconnection of magnetic fields (LV99).
This perpendicular turbulent mixing 
results in the scale-dependent anisotropy of MHD turbulence and 
makes the essential distinction between MHD and hydrodynamic turbulence.

The turbulent energy injected at a large scale, i.e., the injection scale $L$, 
progressively cascades down to small scales. 
The nonlinear cascade rate of Alfv\'{e}nic turbulence, i.e., the eddy-turnover rate, 
determines the perpendicular turbulent mixing rate, 
\begin{equation}\label{eq: velgs}
  \tau_\text{cas}^{-1} = v_l l_\perp^{-1} = V_\text{st} L_\text{st}^{-\frac{1}{3}}  l_\perp^{-\frac{2}{3}}  ,
\end{equation}
with the local turbulent velocity $v_l$ at the length scale $l_\perp$ measured in the direction 
perpendicular to the {\it local} magnetic field. 
The critical balance between $\tau_\text{cas}^{-1}$ and 
the frequency of Alfv\'{e}n waves propagating along the magnetic field is satisfied in the strong MHD turbulence regime 
(GS95).
In the above expression, the GS95 scaling for the strong MHD turbulence is used: 
\begin{equation}\label{eq: gsloc}
   v_l = V_\text{st} (l_\perp / L_\text{st})^{\frac{1}{3}} , 
\end{equation}
where $V_\text{st}$ is the turbulent velocity at the injection scale $L_\text{st}$ of strong MHD turbulence. 
More specifically, in super-Alfv\'{e}nic turbulence with dominant turbulent kinetic energy at $L$, there is 
\begin{equation}\label{eq: supalf}
   V_\text{st} = V_\mathrm{A},  L_\text{st} = L \mathcal{M}_\mathrm{A}^{-3} , 
\end{equation}
where $V_\mathrm{A}$ is the Alfv\'{e}n speed, 
$\mathcal{M}_\mathrm{A} = V_L / V_\mathrm{A} >1$ is the Alfv\'{e}n Mach number, and $V_L$ is the turbulent velocity at $L$. 
Within the range $[L, L_\text{st}]$, the turbulence is in the hydrodynamic regime with isotropic turbulent mixing 
and isotropic scaling. 
In the case of sub-Alfv\'{e}nic turbulence with dominant magnetic energy at $L$, i.e., $\mathcal{M}_\mathrm{A} <1$, there is
\citep{Lazarian06}, 
\begin{equation}\label{eq: subalf}
   V_\text{st} = V_L \mathcal{M}_\mathrm{A}, L_\text{st} = L \mathcal{M}_\mathrm{A}^2   .
\end{equation}
Within $[L, L_\text{st}]$, it is the weak turbulence resulting from weakly interacting Alfv\'{e}n waves
(LV99).

Given the expression of the perpendicular mixing rate (Eq. \eqref{eq: velgs}), the critical balance condition
\begin{equation}\label{eq: criba}
    \tau_\text{cas}^{-1} = V_\mathrm{A} / l_\|
\end{equation}
leads to the anisotropic scaling relation of strong MHD turbulence, 
\begin{equation}\label{eq: turani}
   l_\| = \frac{V_\mathrm{A}}{V_\text{st}} L_\text{st}^{\frac{1}{3}} l_\perp^{\frac{2}{3}},
\end{equation}
where $l_\|$ is length scale measured parallel to the magnetic field.
It shows that the anisotropy of strong MHD turbulence is scale-dependent, with 
smaller-scale turbulent eddies more elongated along the magnetic field.

As mentioned above, slow modes that induce density fluctuations do not affect the dynamics of MHD turbulence. 
The perpendicular mixing of density fluctuations by Alfv\'{e}n modes is 
analogous to the mixing of a passive scalar (LG01).
The resulting density fluctuations 
follow the same cascade and conform to the same anisotropic scaling as Alfv\'{e}nic turbulent velocities. 
Shaped by the turbulence anisotropy, 
density fluctuations tend to exhibit elongated structures along the magnetic field, i.e., parallel density filaments. 

After the injection of a density source $\delta n(l_0)$, 
as the turbulent mixing operates toward the homogenization of density variations, 
lower-density fluctuations develop on lengths scales below the density injection scale $l_0$ with the scaling 
\begin{equation}
    \delta n (l_\perp) = \delta n(l_0) (l_\perp / l_0 )^\frac{1}{3}.
\end{equation}
Thus we expect that parallel filaments formed in MHD turbulence tend to have a relatively low density contrast.
Following the turbulent energy cascade,
the lifetime of a parallel filament is determined by the turnover time of the turbulent eddy at the perpendicular length scale equal 
to the filament width.

\subsection{Numerical tests on the formation mechanism of low-density parallel filaments}

We perform 3D simulations of driven MHD turbulence as numerical tests for the above theory. The FLASH code \citep{fryxell00} is deployed, with a directionally unsplit staggered mesh (USM) MHD solver based on a finite-volume, high-order Godunov method combined with a constrained transport (CT) type scheme \citep{tzeferacosetal12, lee2013solution}, to solve the equations of inviscid ideal magnetohydrodynamics as follows:
  \begin{subequations}
    \begin{gather}
      \frac{\partial\rho}{\partial t} + \bm{\nabla} \cdot (\rho \bm {v}) = 0 \\
      \frac{\partial\rho \bm{v}}{\partial t} + \bm{\nabla} \cdot (\rho \bm{v} \bm{v} - \bm{B} \bm{B}) + \bm{\nabla} p_* = \bm{s}_\mathrm{stir} \\
      \frac{\partial \bm{B}}{\partial t} + \bm{\nabla} \cdot (\bm{v}\bm{B} - \bm{B} \bm{v}) = 0 \\
      \nabla \cdot \bm{B} = 0
    \end{gather}
    \label{eq:MHD}
  \end{subequations}
  where $p_* = p + B^2 / (8 \pi)$ is the total pressure including both gas pressure $p$ and magnetic pressure $B^2 / (8 \pi)$, and $\bm{s}_\mathrm{stir}$ is the source term of stirring. Isothermal condition is applied during the evolution of these governing equations.

 Our simulations are performed in a 3D Cartesian box with the size of $L_\mathrm{box} = 10$ along each dimension and the resolutions of $128^3$, $256^3$ and $512^3$. To examine the effect of turbulent mixing on density fluctuations, we adopt the solenoidal driving of turbulence, which does not generate density structures associated with compressible modes. The stirring is performed over wavenumbers of $1 \leq n \leq 4$, which corresponds to the $k$ vector of $0.1 \leq k\equiv n/L_\mathrm{box} \leq 0.4$ and the length scale of $2.5 \leq l\equiv k^{-1} \leq 10$, with a driving amplitude of paraboloidal shape in Fourier space and an autocorrelation time of $t_\mathrm{correlation} \equiv L_\mathrm{box} / 2 V = 5$, where $V = 1$ is the RMS velocity when the simulations reach a statistically stable state \citep{federrath2010comparing}. The simulations are initialized with a uniform density and sound speed, and constant magnetic fields along x-axis. These initial values are carefully chosen in order that after stirring over a certain period ($\sim$ a few largest eddy turnover time $t_{\mathrm{eddy},L} = L / V$), the simulations become statistically stable at $\mathcal{M}_\mathrm{s} \sim 0.5$ and $\mathcal{M}_\mathrm{A} \sim 1$, where the sonic- and Alfv\'{e}n--Mach numbers are RMS values averaged over the entire box. These values are typical for the subsonic diffuse warm medium with comparable turbulent and magnetic energies at the driving scale of turbulence 
\citep{Beck96}.

  \begin{figure}
    \begin{center}
      \includegraphics[width=0.5\textwidth]{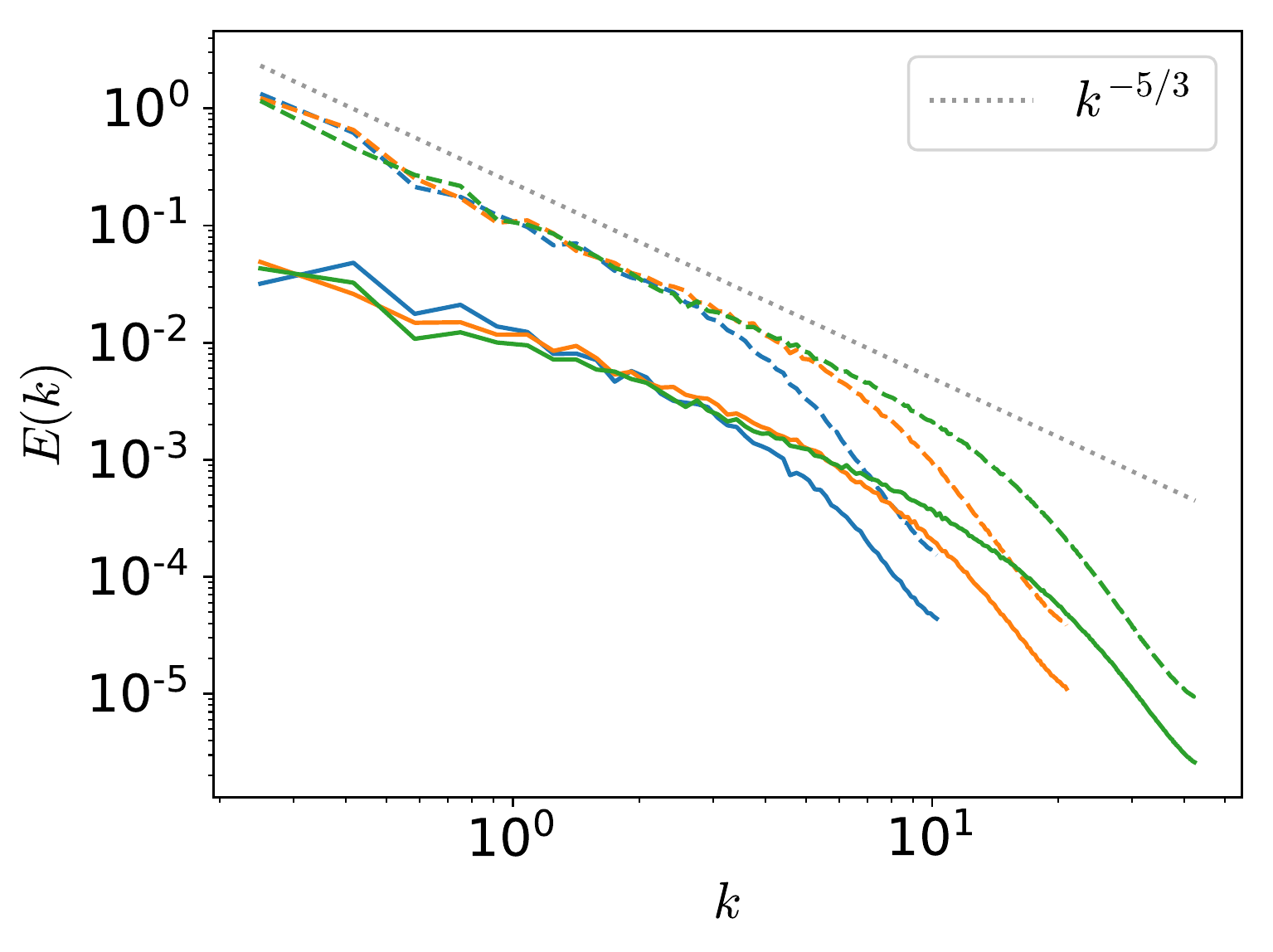}
      \caption{Energy power spectra of density fluctuations (solid lines) and velocity fluctuations (dashed lines) from snapshots at $t\sim 6\ t_{\mathrm{eddy},L}$ with different resolutions: blue --- resolution of $128^3$, orange --- resolution of $256^3$, green --- resolution of $512^3$. The slope of Kolmogorov spectrum
      $k^{-5/3}$ is indicated by the dotted gray line.}
      \label{fig:spectrum}
    \end{center}
  \end{figure}

Fig. \ref{fig:spectrum} presents the energy spectra of both density fluctuations $\rho/\bar{\rho}$ (density normalized by mean density $\bar{\rho}$) and velocity fluctuations 
(velocity magnitude normalized by sound speed), from the snapshot of $t \sim 6\ t_{\mathrm{eddy},L}$. As expected, density fluctuations passively follow the 
same cascade as turbulent velocities, and as a result, 
the density spectrum exhibits the same 
Kolmogorov form as that of the velocity spectrum. In addition, these spectra demonstrate a nice convergence within available inertial range in each of the simulations with different resolutions. Hereinafter, we focus on the simulations with the highest resolution of $512^3$ for our analysis.

  \begin{figure*}
    \subfloat[Slice plots of density fluctuations at different times, viewed along z-axis (upper panel) and x-axis (lower panel).]{
      \begin{tabular}{c}
        \includegraphics[width=\textwidth]{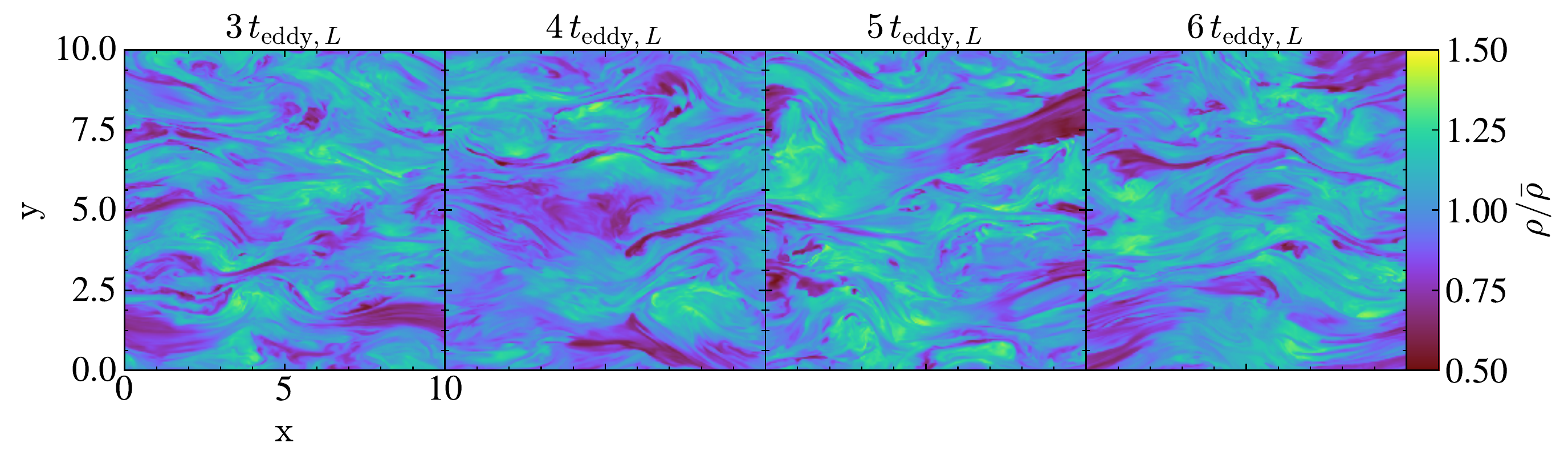} \\
        \includegraphics[width=\textwidth]{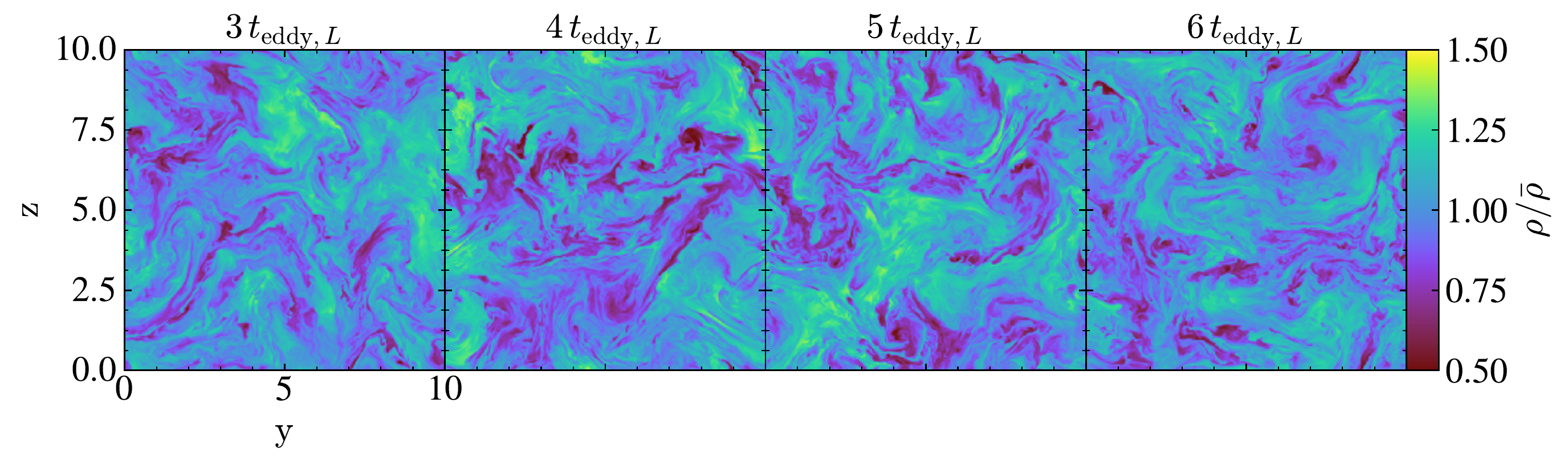}
      \end{tabular}
      \label{fig:dens_t}
      }
    \\
    \subfloat[Slice plots of density fluctuations superposed with magnetic field lines at $t=6\ t_{\mathrm{eddy},L}$, viewed along z-axis (left panel) and x-axis (right panel).]{
      \includegraphics[width=0.5\textwidth]{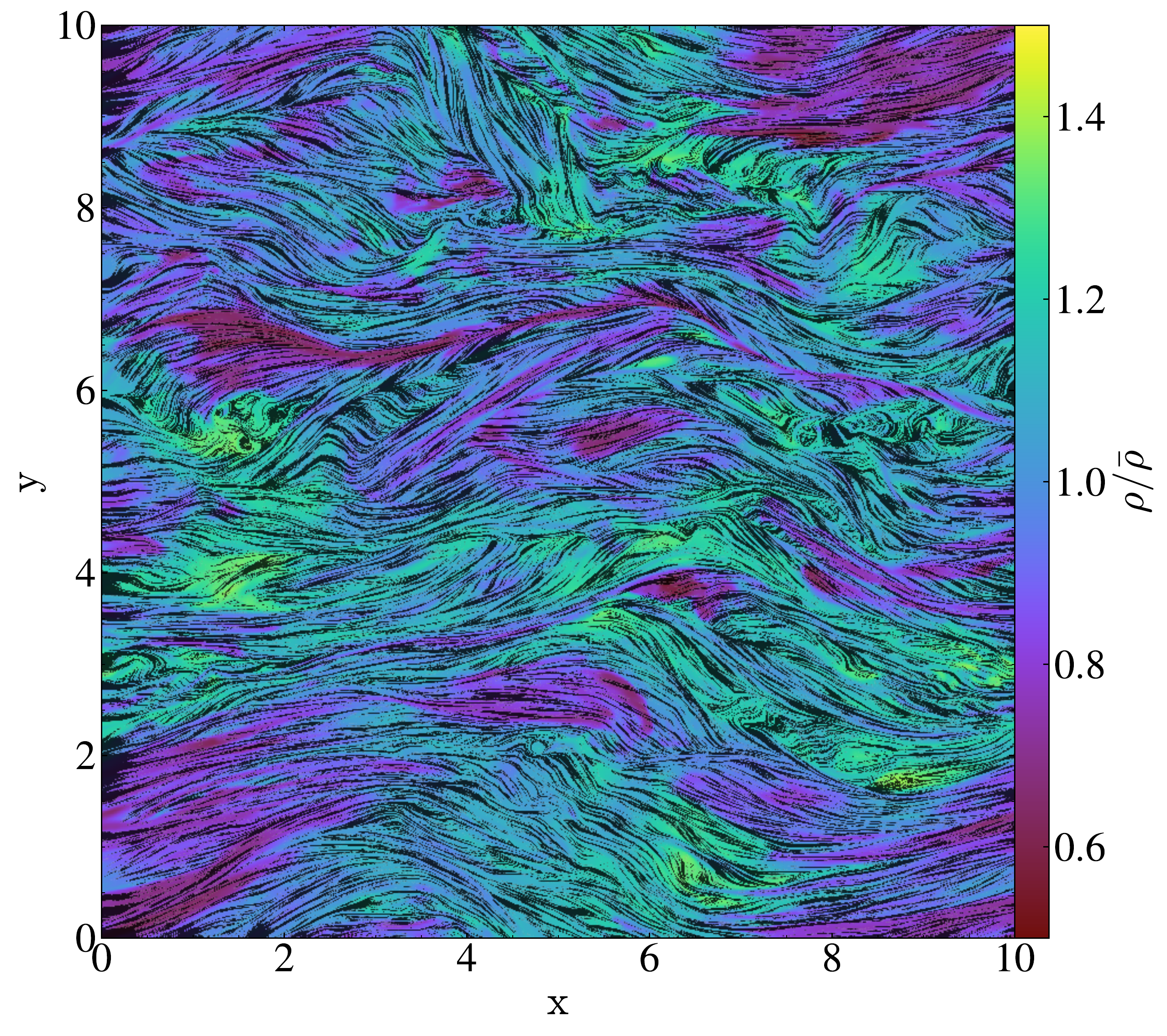}
      \includegraphics[width=0.5\textwidth]{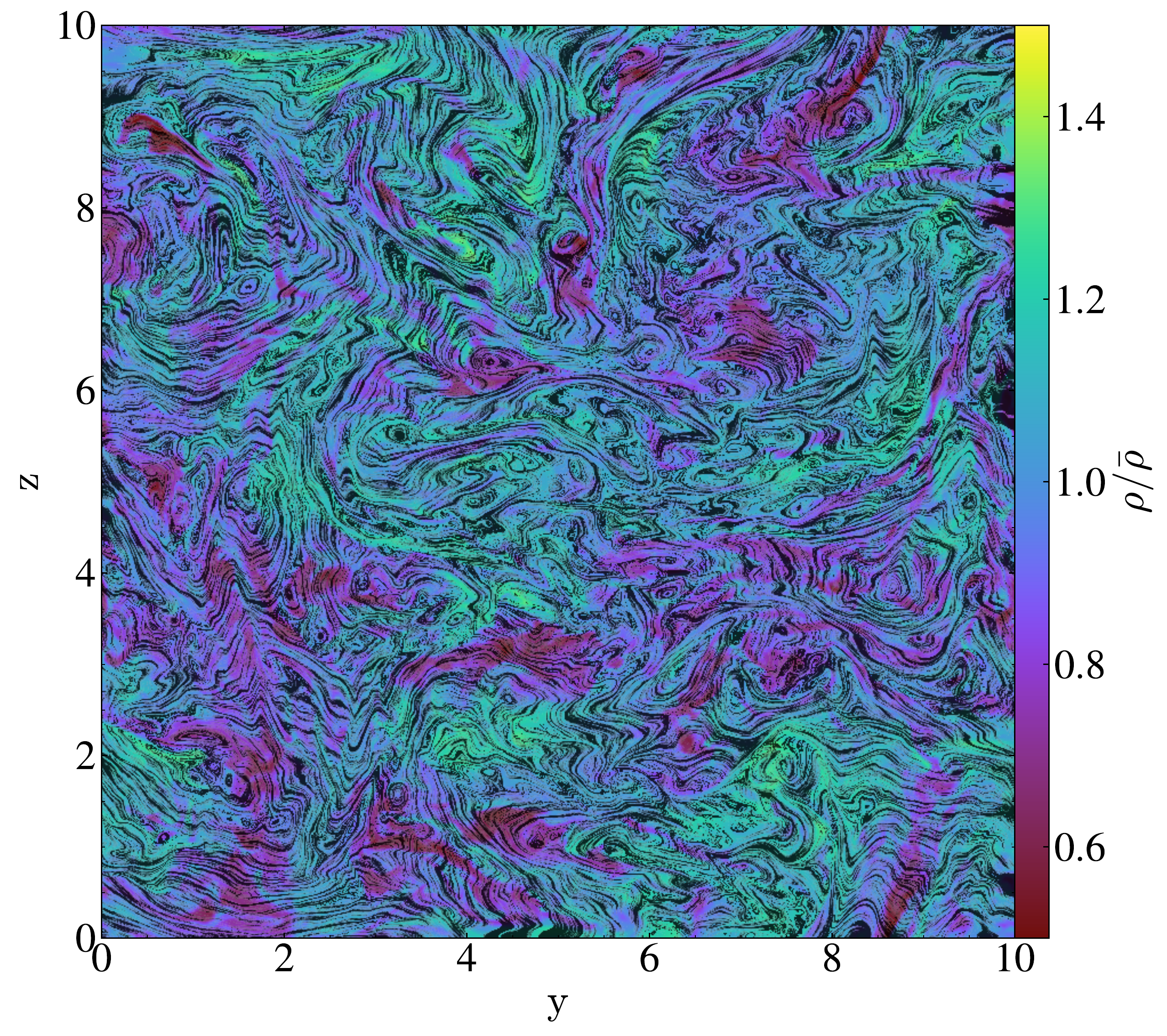}
      \label{fig:dens_B}
      }
    \caption{The structures of density fluctuations and magnetic fields.}
    \label{fig:dens}
  \end{figure*}

  \begin{figure*}
    \subfloat[Slice plots of dyed density fluctuations at different times, viewed along z-axis (upper panel) and x-axis (lower panel).]{
      \begin{tabular}{c}
        \includegraphics[width=\textwidth]{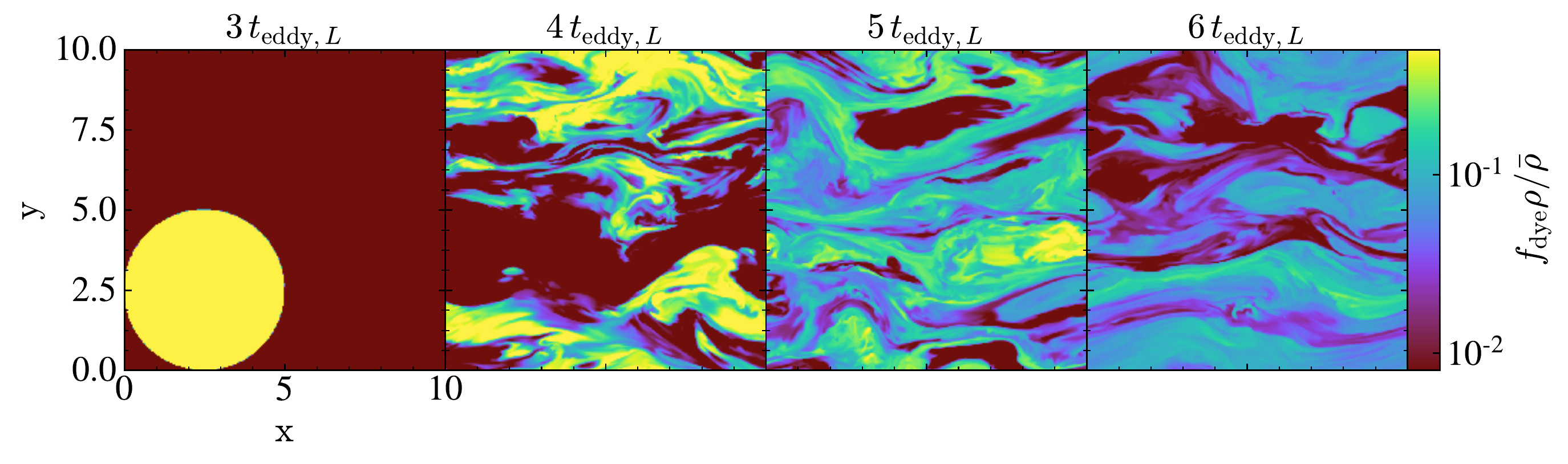} \\
        \includegraphics[width=\textwidth]{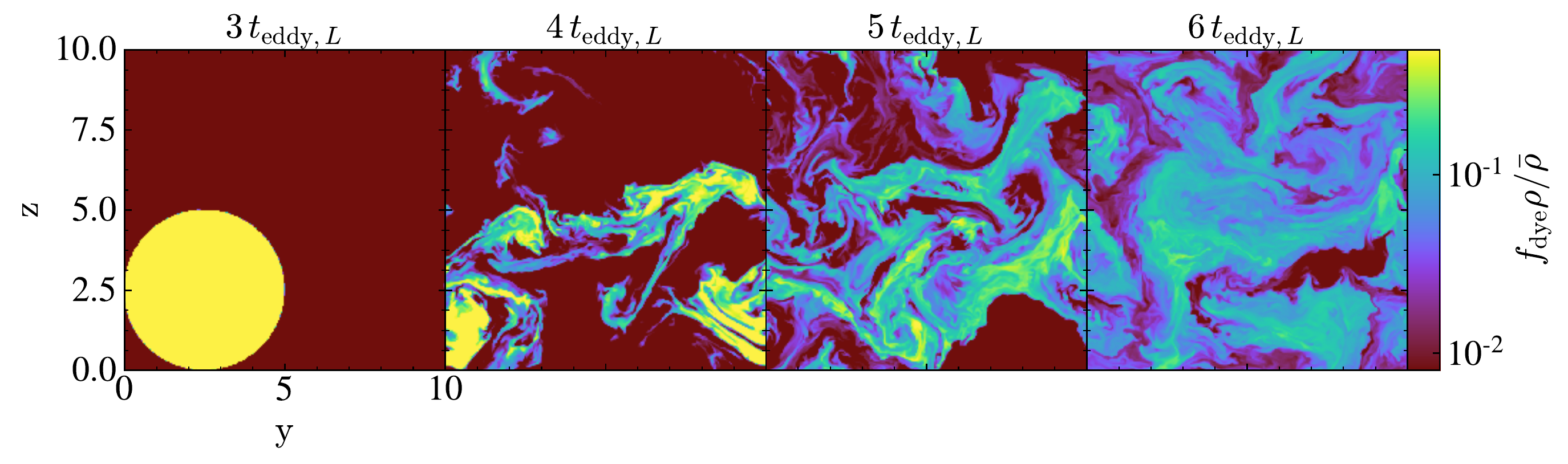}
      \end{tabular}
      \label{fig:dye_t}
      }
    \\
    \subfloat[Slice plots of dyed density fluctuations superposed with magnetic field lines at $t=6\ t_{\mathrm{eddy},L}$, viewed along z-axis (left panel) and x-axis (right panel).]{
      \includegraphics[width=0.5\textwidth]{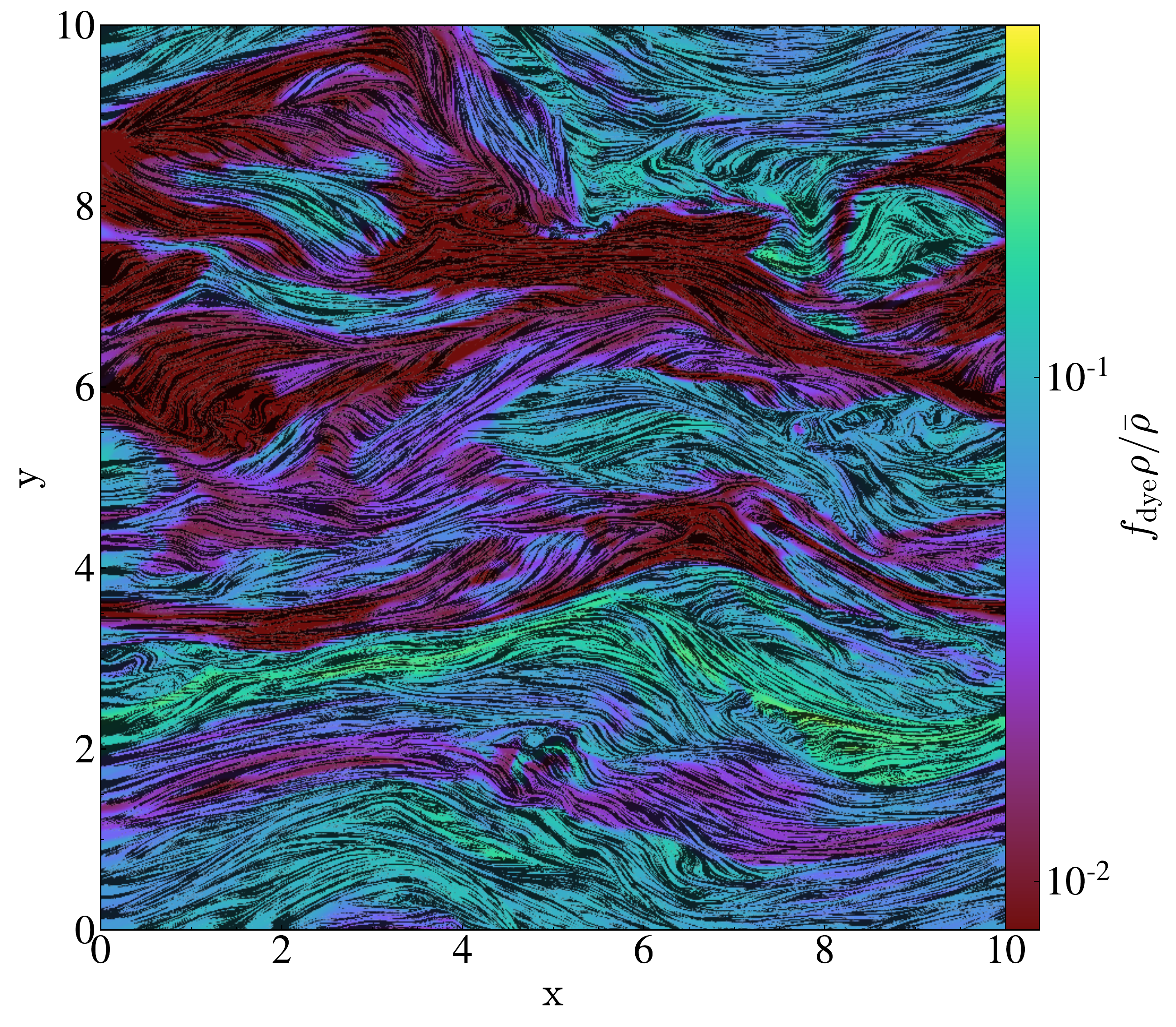}
      \includegraphics[width=0.5\textwidth]{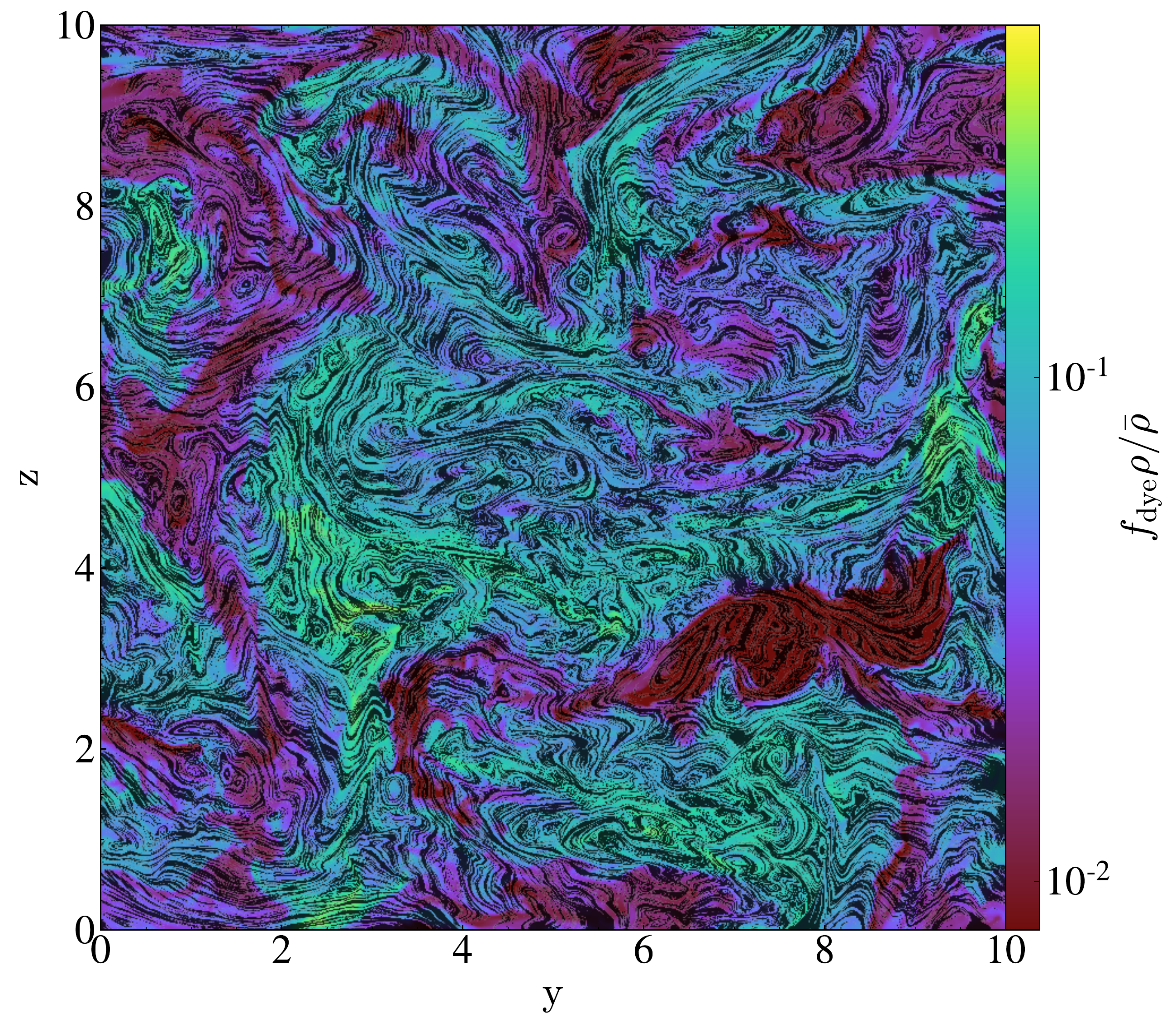}
      \label{fig:dye_B}
      }
    \caption{The structures of dyed density fluctuations and magnetic fields.}
    \label{fig:dye}
  \end{figure*}

To numerically test the perpendicular turbulent mixing of density fluctuations, at $t = 3\ t_{\mathrm{eddy},L}$ when the simulation becomes statistically stable, we assign certain regions with a dimensionless ``dye'' value of $f_\mathrm{dye} = 1$, which is a mass scalar passively advected with density: $\partial_t \left(f_\mathrm{dye} \rho\right) + \nabla \cdot (f_\mathrm{dye} \rho \bm{v}) = 0$,
and trace the time evolution of its spatial configuration. In Fig. \ref{fig:dens} and Fig. \ref{fig:dye}, we show the slice plots of 
the gas density fluctuations $\rho/\bar{\rho}$ and dyed density fluctuations $f_\mathrm{dye} \rho / \bar{\rho}$ at different times,
as well as the magnetic field geometries, viewed along z-axis and x-axis respectively. 
Note that the magnetic field lines are initialized aligned along x-axis.
When the simulation is statistically stable, 
we find $\langle |\delta B| \rangle / \bar{B} \sim 0.7$, where $\delta B$ is the magnetic field perturbation, $\bar{B}$ is the mean magnetic field, 
and $\langle ... \rangle$ denotes an ensemble average.
In a situation with $\langle |\delta B| \rangle / \bar{B} > 1$, we do not expect to see significantly different 
magnetic field and density structures at different line-of-sight orientations.

In Fig. \ref{fig:dye}, from the snapshots of density fluctuations taken at different times 
we see that 
under the effect of perpendicular turbulent mixing, 
the initial dyed density fluctuations with an isotropic spatial distribution evolves into 
smaller-scale anisotropic density structures with lower density variations. 
The resulting small-scale density structures are highly elongated along the local magnetic field direction.

\begin{figure*}
      \begin{tabular}{cc}
        \includegraphics[width=0.5\textwidth]{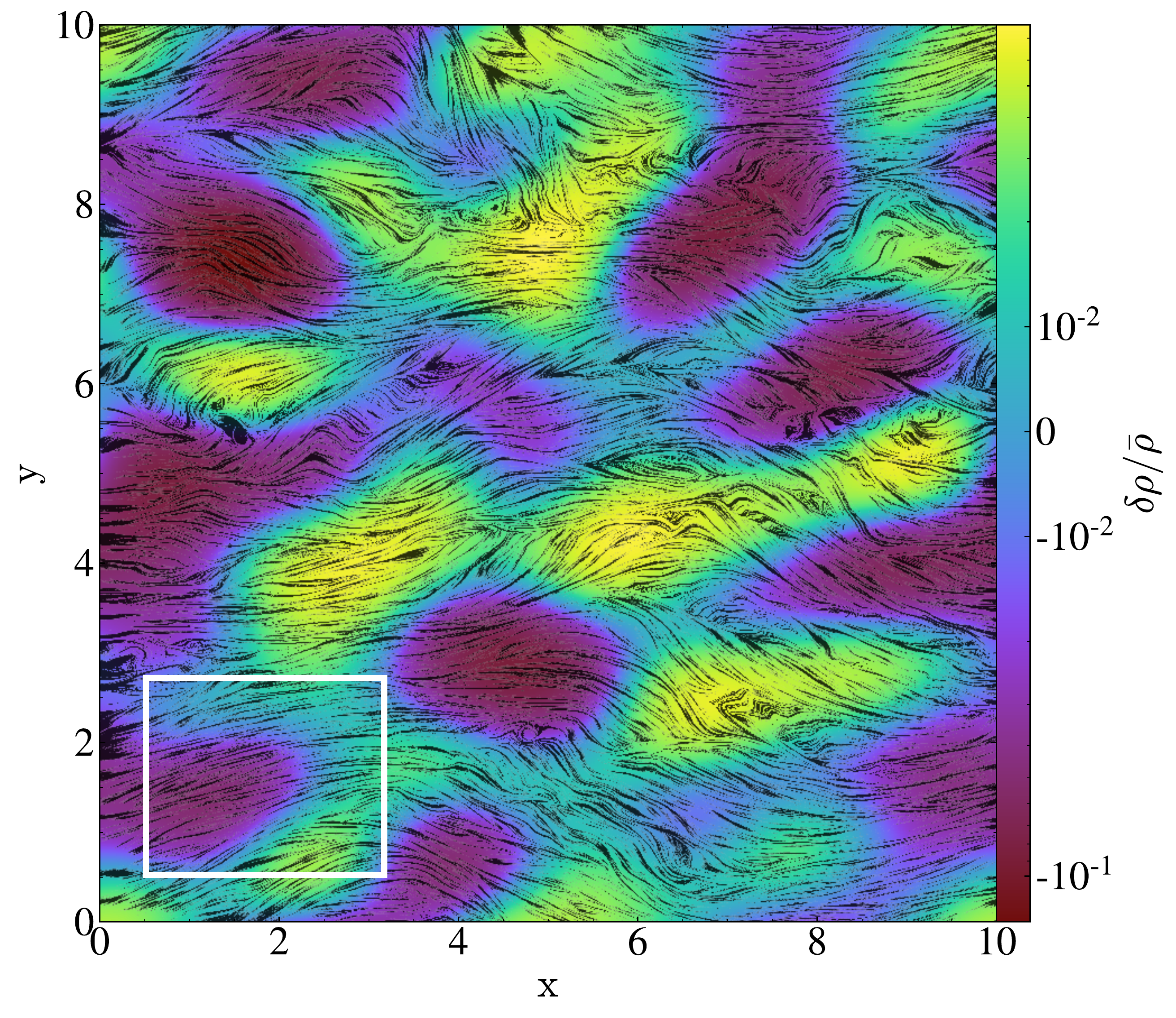} &
        \includegraphics[width=0.5\textwidth]{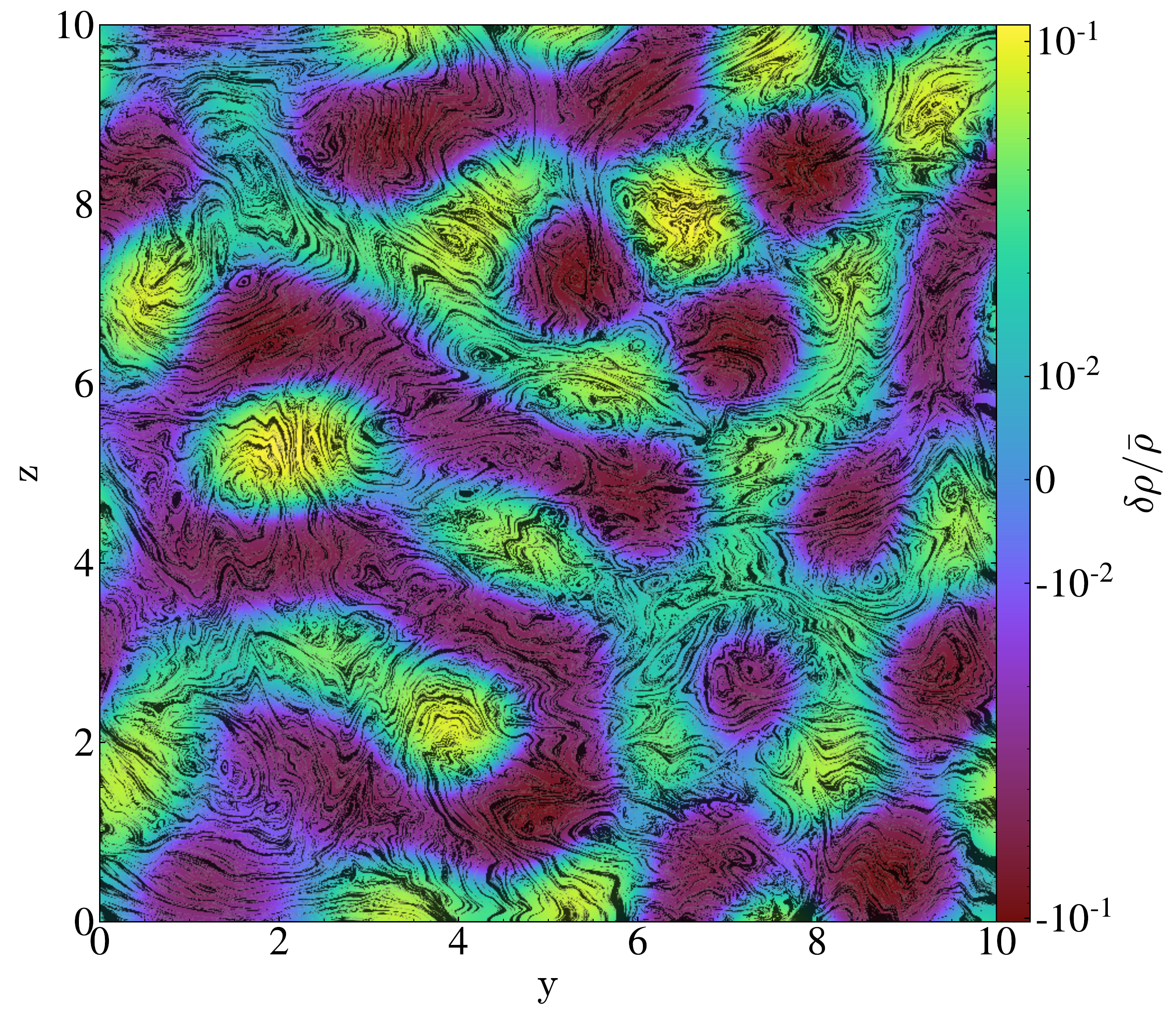} \\
        \includegraphics[width=0.5\textwidth]{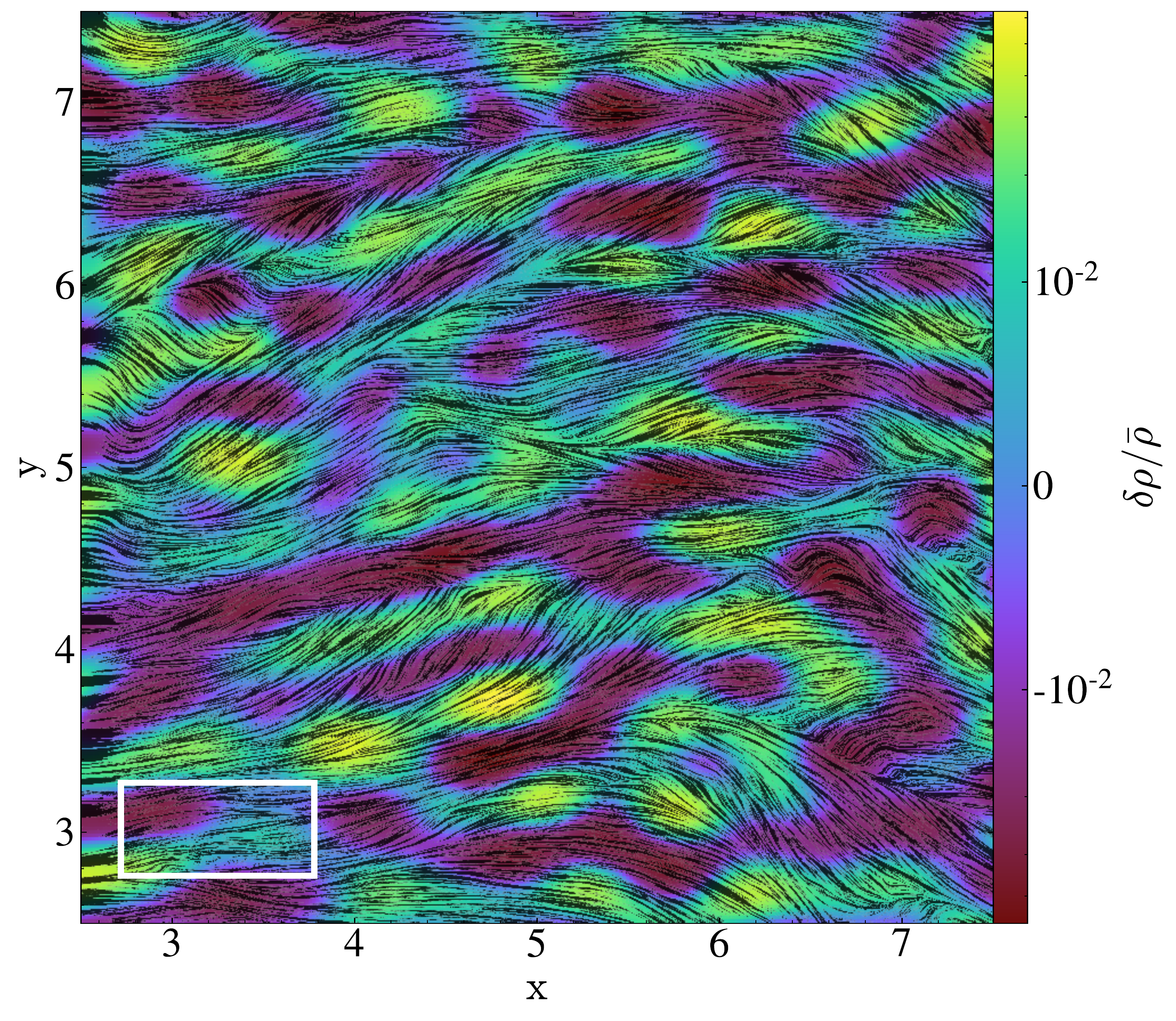} &
        \includegraphics[width=0.5\textwidth]{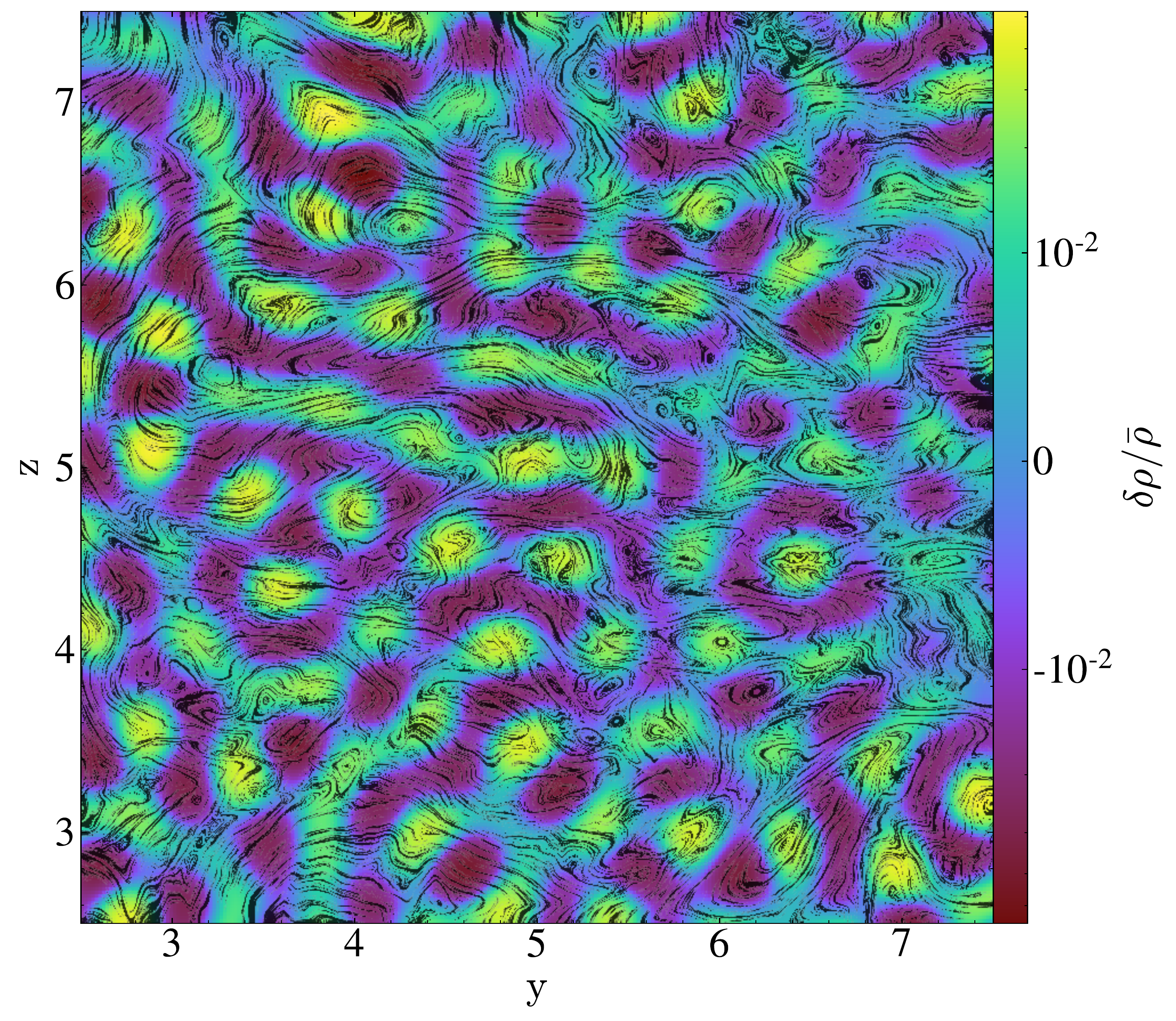} \\
        \includegraphics[width=0.5\textwidth]{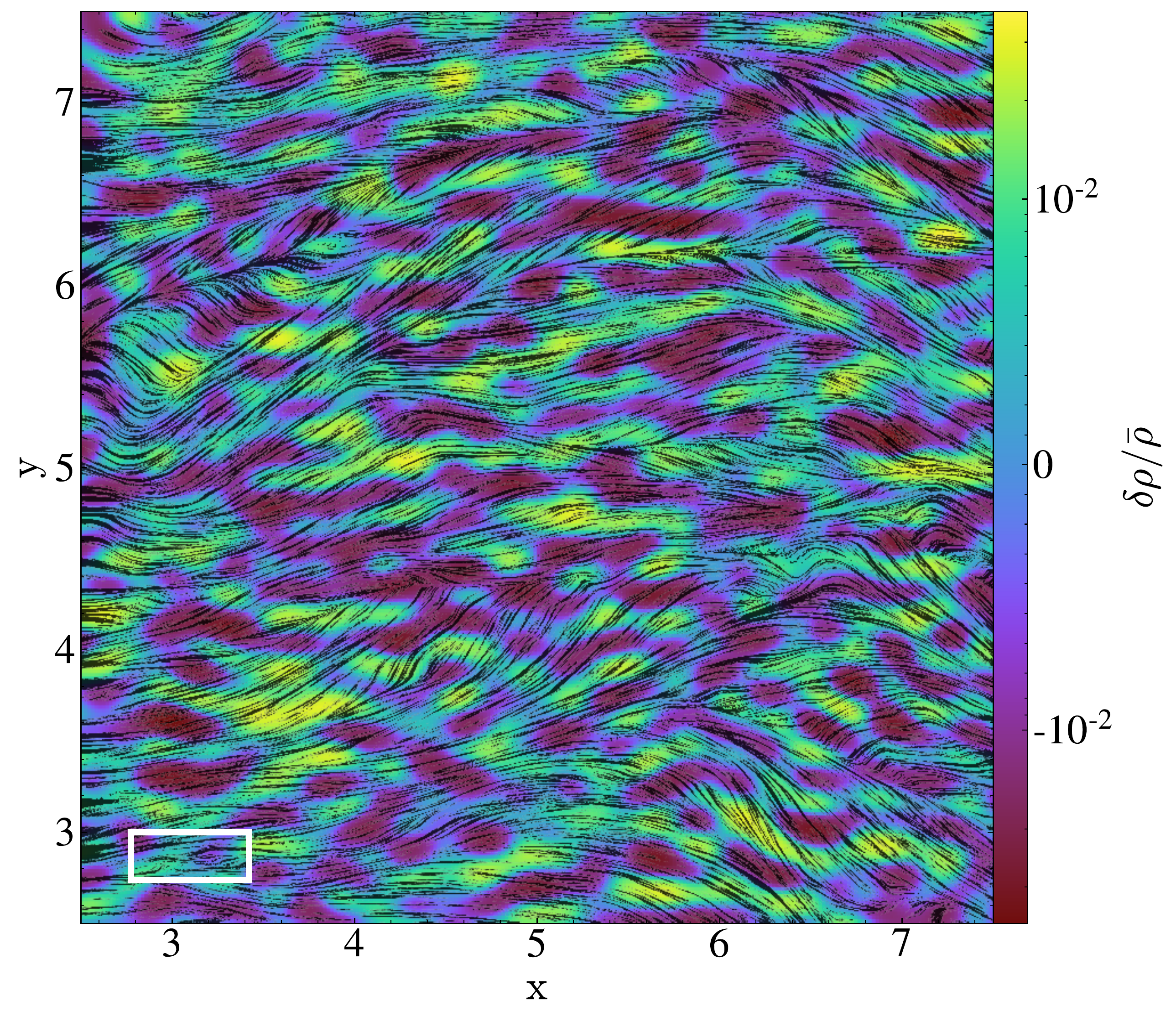} &
        \includegraphics[width=0.5\textwidth]{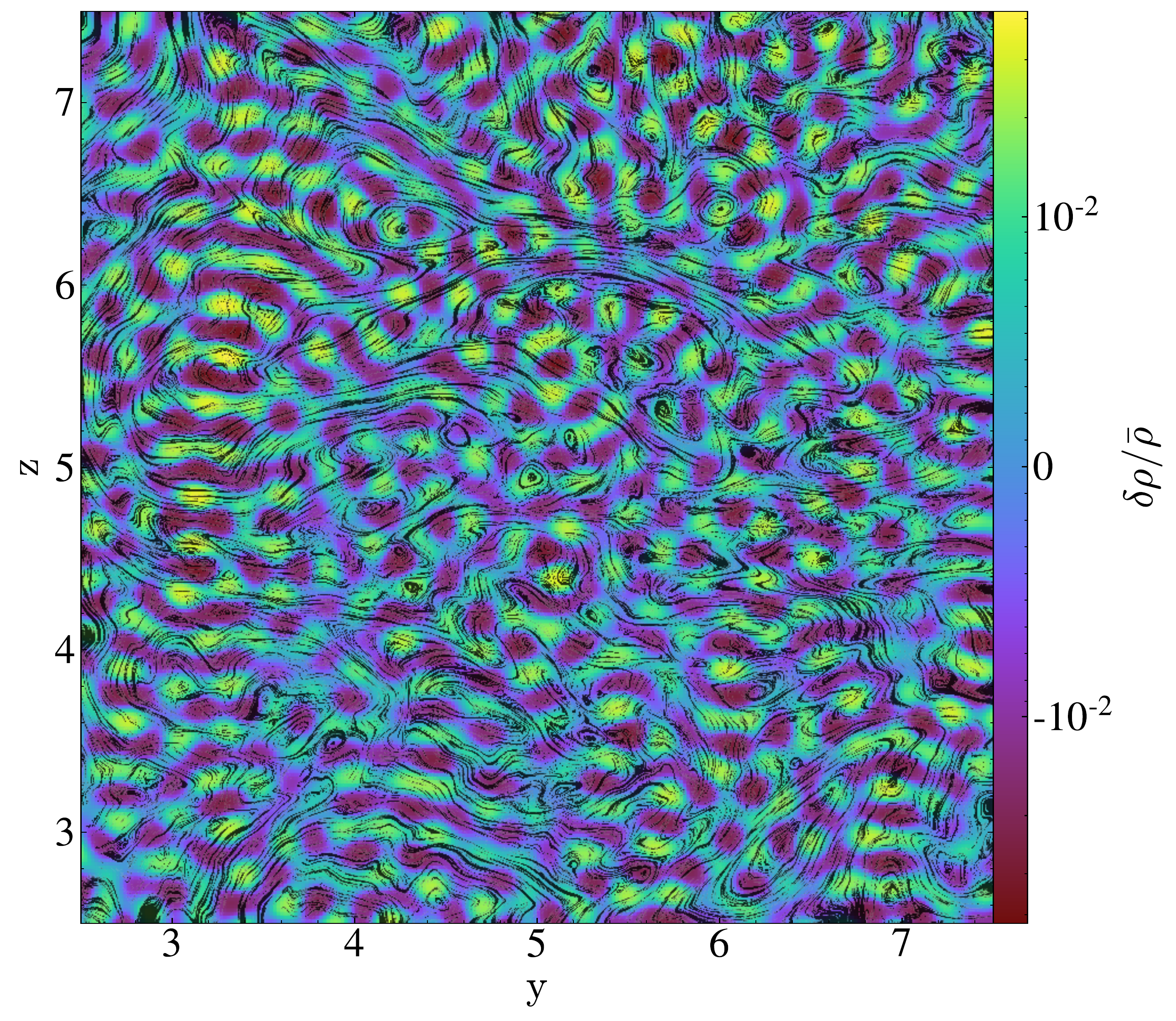}
      \end{tabular}
 \caption{Slice plots of density fluctuations at $t=6\ t_{\mathrm{eddy},L}$, superposed with magnetic field lines: 
upper panel: $0.45< k < 0.55$, 
middle panel: $1.95 < k < 2.05$, 
lower panel: $3.95< k <4.05$,
left panel: viewed along z-axis,
right panel: viewed along x-axis. 
The figures in the middle and lower panel are zoomed in by a factor of $2$.
The rectangles giving the theoretical predictions (Eq. \eqref{eq: turani})
for the anisotropic scaling of density structures at $k=0.45$, $1.95$, and $3.95$ are overplotted.}     
      \label{fig:dens_scale}
  \end{figure*}

\begin{figure*}
      \begin{tabular}{cc}
        \includegraphics[width=0.5\textwidth]{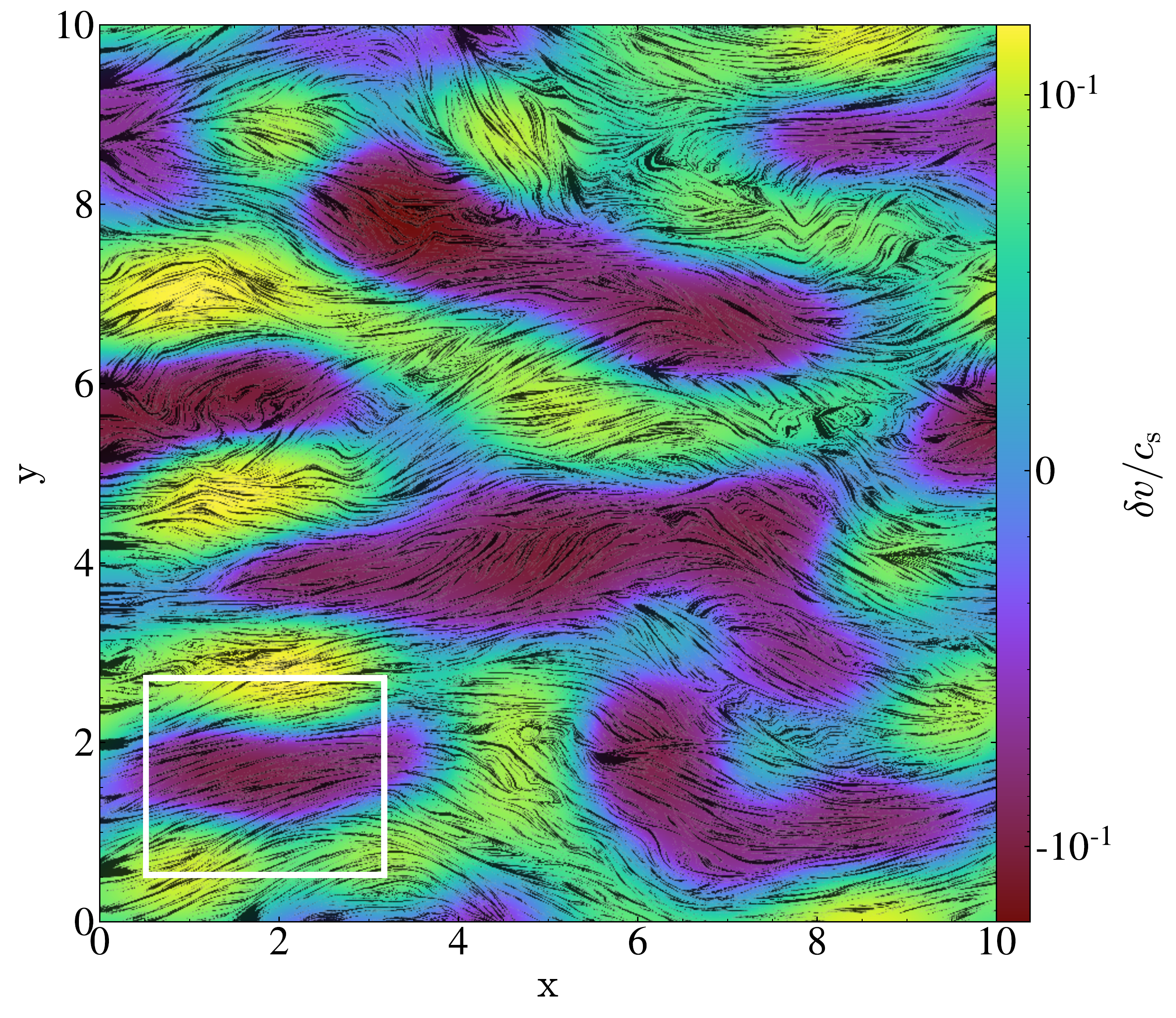} &
        \includegraphics[width=0.5\textwidth]{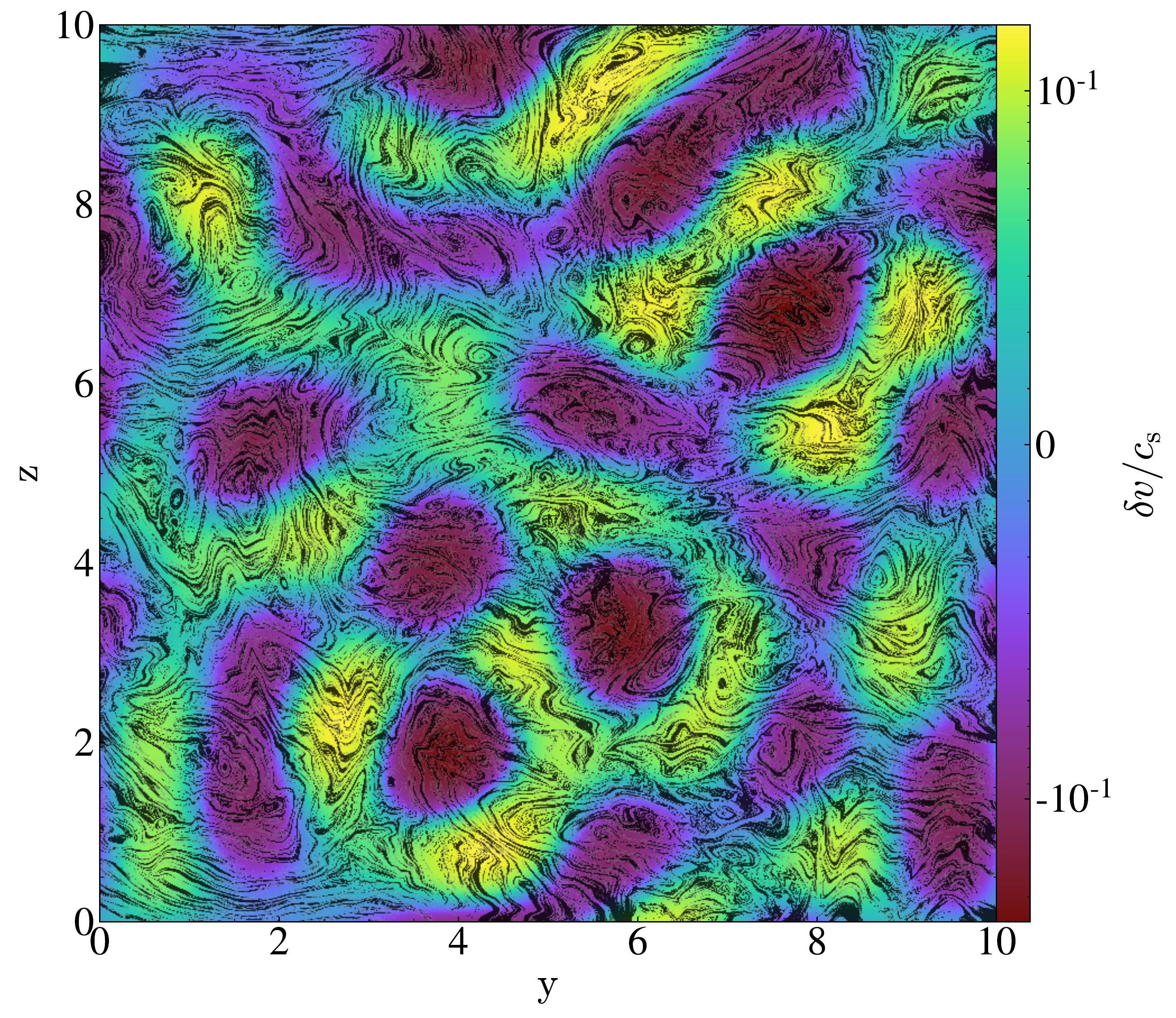} \\
        \includegraphics[width=0.5\textwidth]{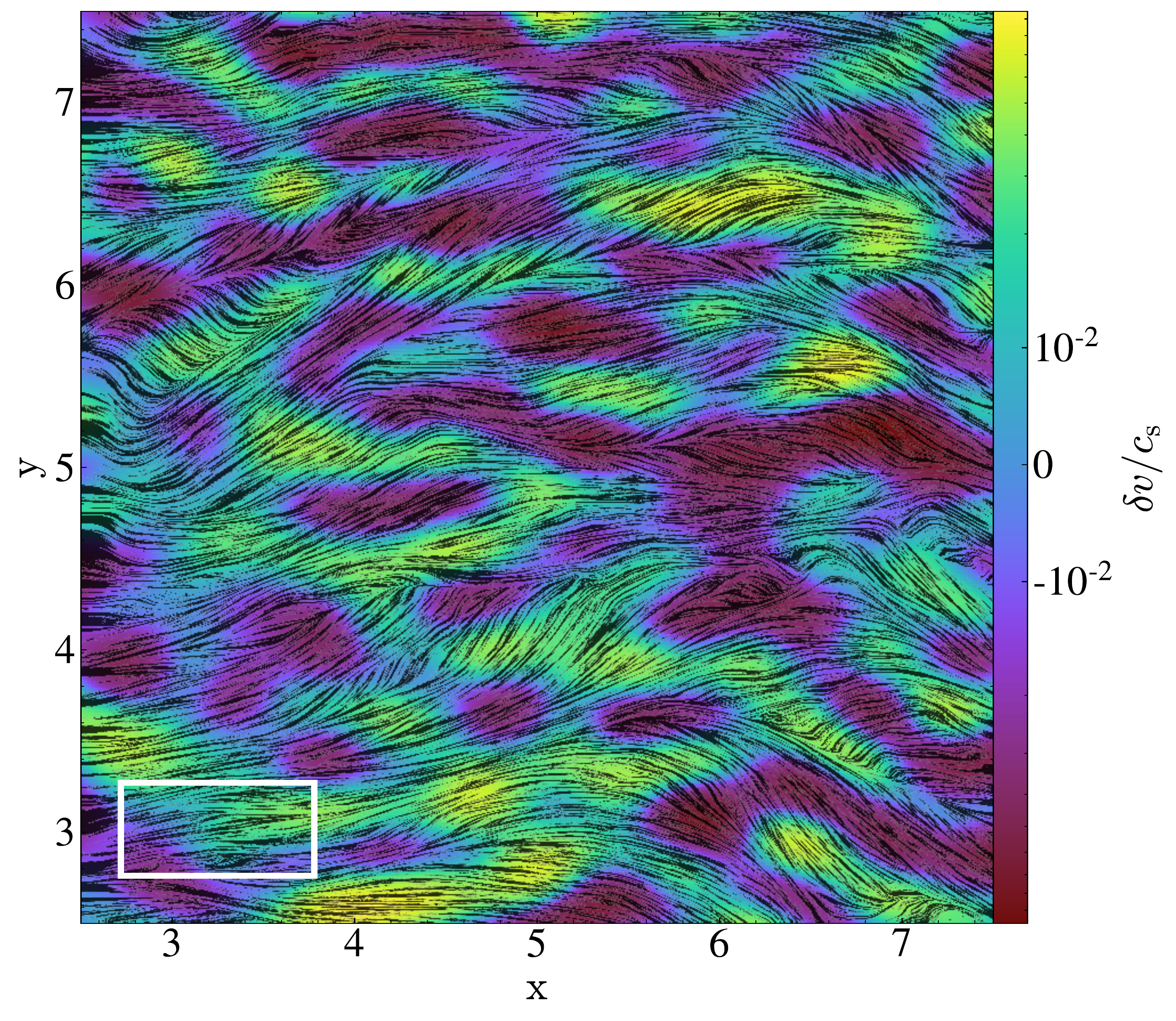} &
        \includegraphics[width=0.5\textwidth]{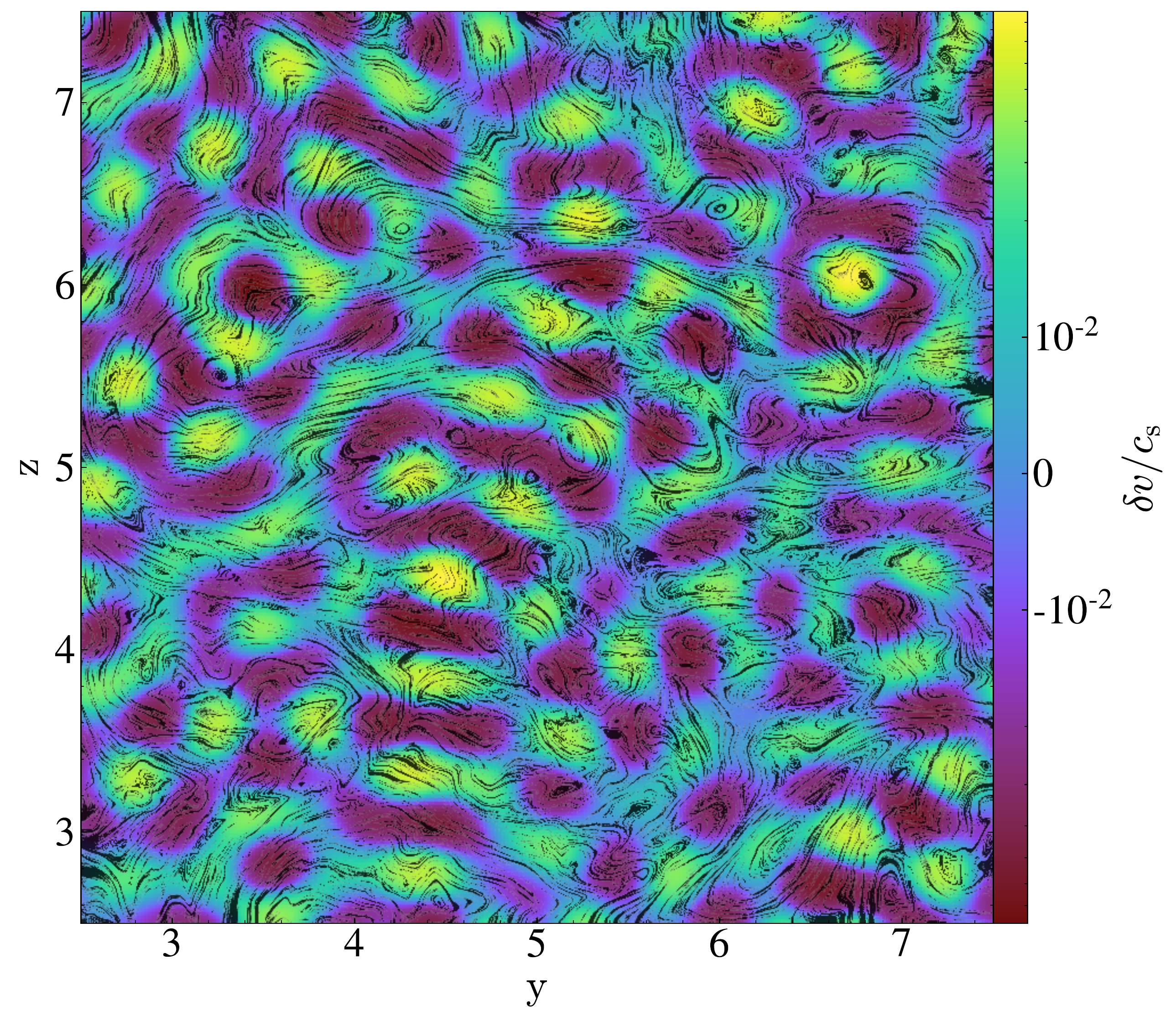} \\
        \includegraphics[width=0.5\textwidth]{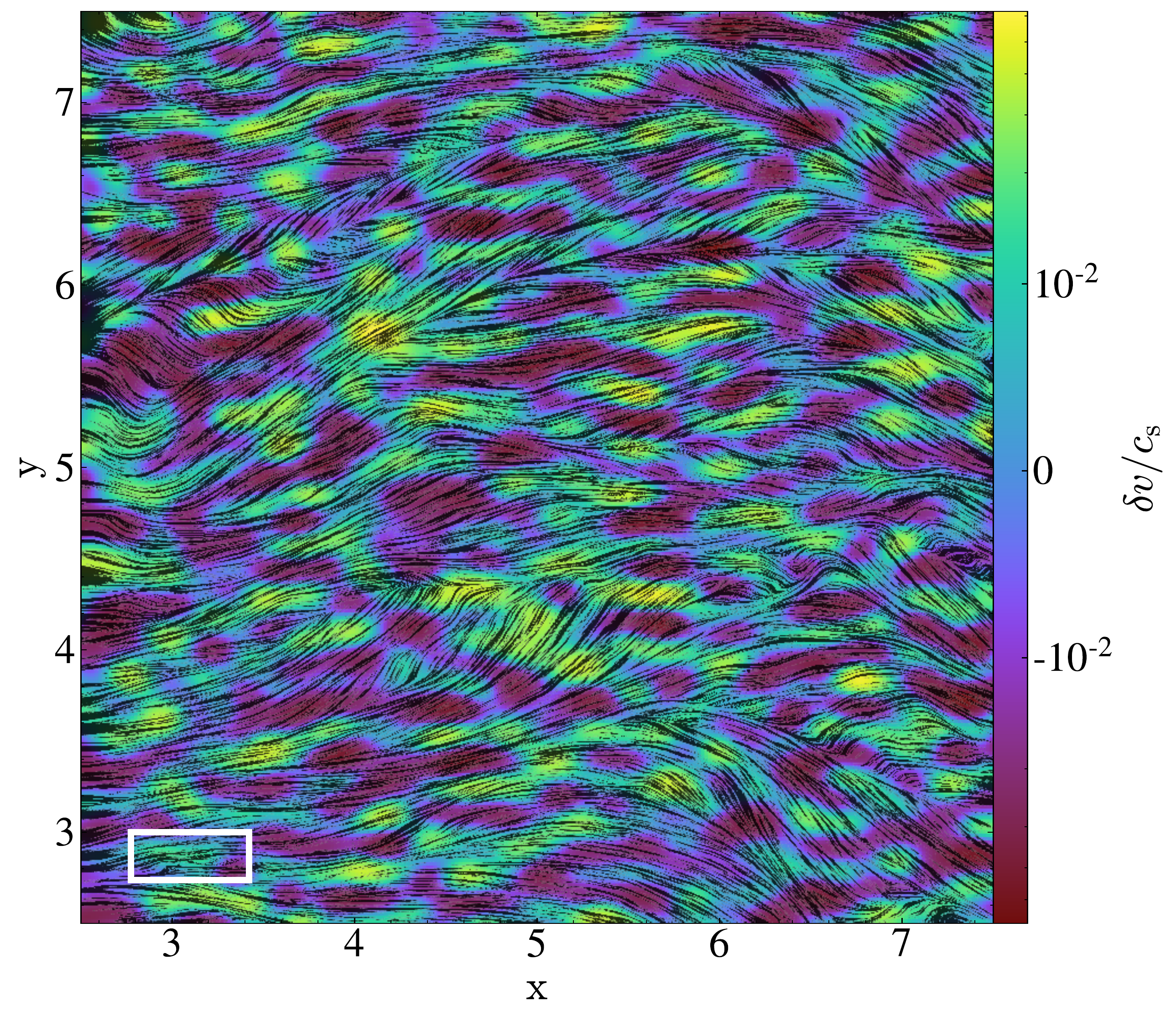} &
        \includegraphics[width=0.5\textwidth]{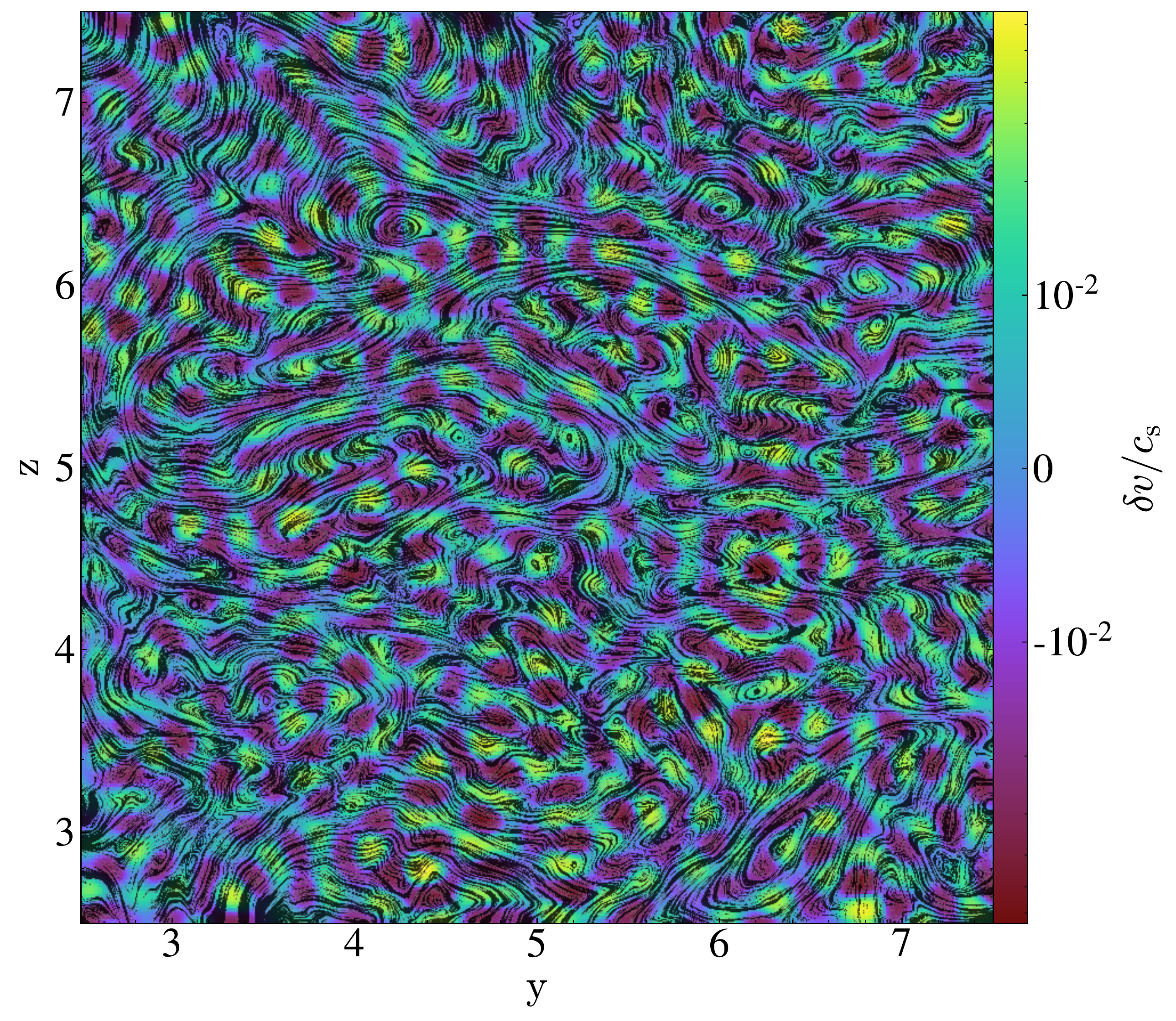} 
      \end{tabular}
 \caption{Same as Fig. \ref{fig:dens_scale} but for velocity fluctuations.}     
      \label{fig:vel_scale}
  \end{figure*}

The perpendicular turbulent mixing leads to the scale-dependent anisotropy of MHD turbulence.
In Fig. \ref{fig:dens_scale}, we display the slice plots of density fluctuations measured within different ranges of length scales. 
To quantify the scale-dependent anisotropy, we overplot rectangles indicating the scaling relation in Eq. \eqref{eq: turani} 
at $k=0.45$, $1.95$, and $3.95$, 
with $l_\perp \approx k^{-1}$, 
$V_\text{st} = V_A$, and $L_\text{st} = L = 4$ where the energy injection peaks. 
We see that the theoretical scaling well describes the anisotropy of density structures, which becomes stronger toward smaller scales.
By comparing with the similar measurements of turbulent velocities in Fig. \ref{fig:vel_scale}, we see the same 
scale-dependent anisotropy of both velocity and density fluctuations as expected. 
The elongated velocity and density structures are both well aligned with the local magnetic field.

\begin{figure*}
      \begin{tabular}{cc}
        \includegraphics[width=0.5\textwidth]{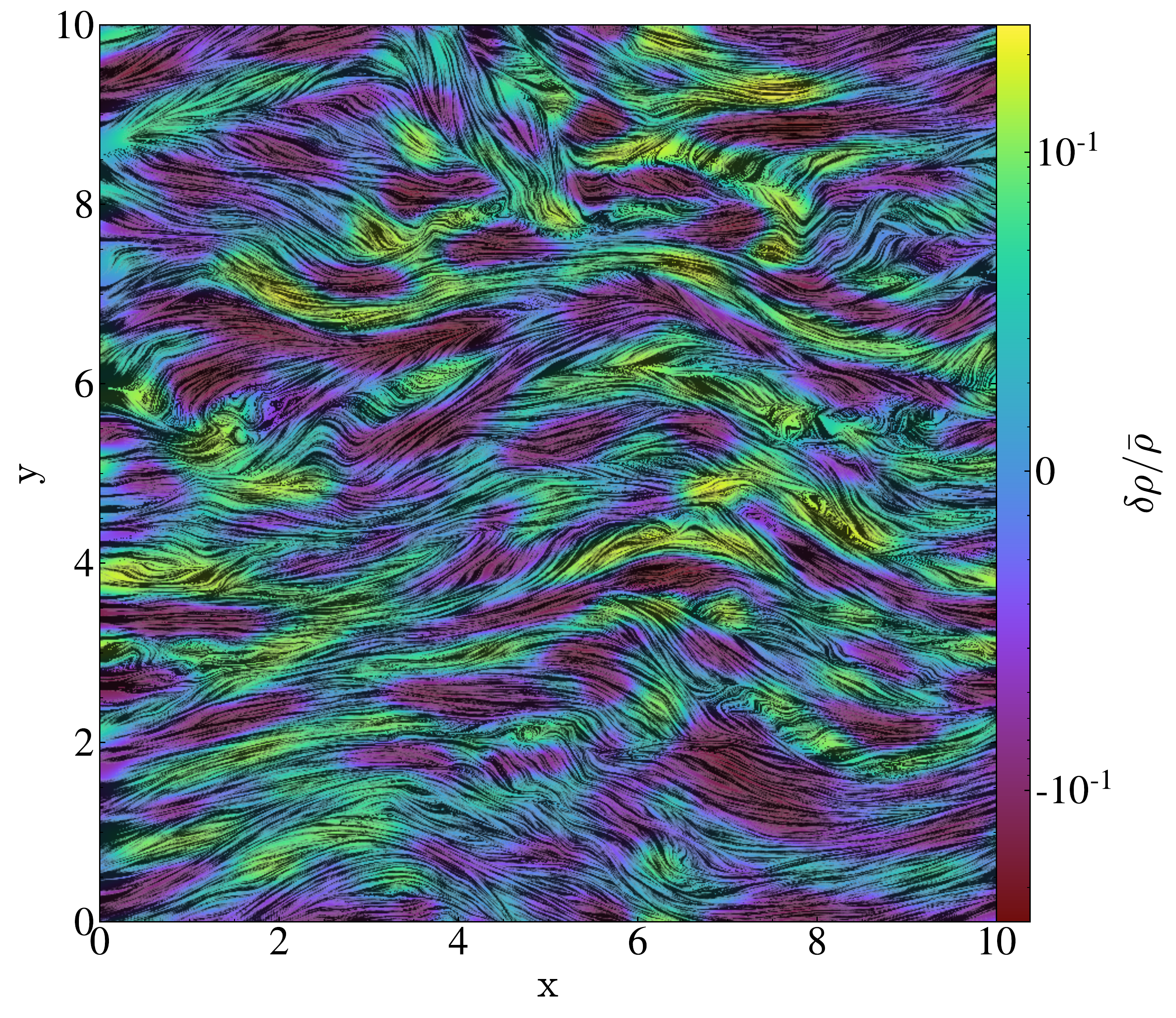} &
        \includegraphics[width=0.5\textwidth]{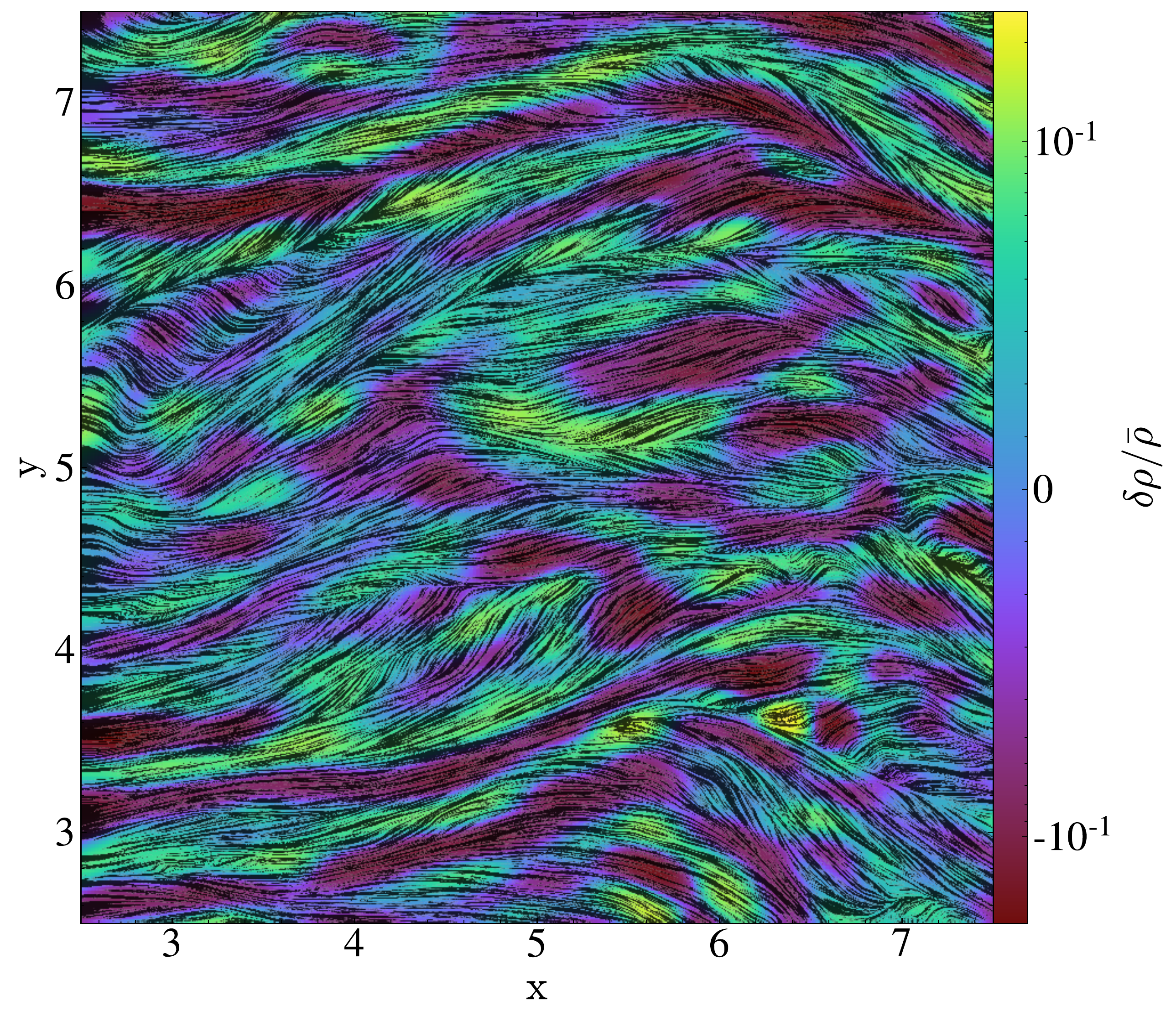} \\
       \end{tabular}
 \caption{Same as Fig. \ref{fig:dens_scale} but for a broader range of $k$:
 left: $1<k<2$, 
 right: $2<k<4$, viewed along z-axis. }     
      \label{fig:denbro}
  \end{figure*}

As the density fluctuations in turbulence tend to spread out in $k$ space (see Fig. \ref{fig:dye}),
we further present the density fluctuations extending over 
a broader range of length scales within the inertial range of turbulence in Fig. \ref{fig:denbro}.
We see that the density structures appear to be more coherent and elongated along magnetic field lines due to the superposition of 
density fluctuations on different length scales. 
Therefore, in a realistic situation involving multi-scale density fluctuations, 
we expect to see more extended filamentary density structures in magnetized turbulence.

The above numerical tests confirm that 
the perpendicular turbulent mixing in MHD turbulence can account for the formation of low-density filaments
parallel to magnetic fields. 
We caution that due to the limited numerical resolution, the inertial range of turbulence in MHD simulations is quite small 
compared with that of the interstellar turbulence 
\citep{Armstrong95,CheL10}.
According to Eq. \eqref{eq: turani}, the degree of turbulence anisotropy depends on the separation between $L_\text{st}$ and $l_\perp$,
\begin{equation}\label{eq: geraas}
   \frac{l_\|}{l_\perp} = \frac{V_\mathrm{A}}{V_\text{st}} \Big(\frac{L_\text{st}}{l_\perp}\Big)^{\frac{1}{3}} . 
\end{equation}
It means that the low-density parallel filaments in the ISM can be much more elongated than those observed in simulations. 
Moreover, as turbulent mixing tends to homogenizes the density distribution, 
multiphase turbulence simulations are necessary for generating prominently visible parallel filaments 
\citep{Kri04}.

\subsection{Low-density parallel filaments in a partially ionized medium}
\label{ssec: niani}

The ISM is commonly partially ionized 
\citep{Draine:2011aa}.
The coupling state between neutrals and ions depends on the length scale of interest. 
If the MHD turbulence cascades down to sufficiently small scales where the two-fluid effect is important, 
we should consider the 
decoupling between neutrals and ions when studying the density structure of neutral gas.

The parallel neutral-ion decoupling scale is defined as 
\begin{equation}\label{eq: panid}
     l_{\text{ni, dec}, \|} = \frac{V_\mathrm{A}}{\nu_\mathrm{ni}}, 
\end{equation}
corresponding to the equalization between the 
Alfv\'{e}n wave frequency and 
the neutral-ion collision frequency $\nu_\mathrm{ni}= \gamma_d \rho_i$, 
where $\gamma_d$ is the drag coefficient 
(see e.g. \citealt{Shu92}), 
and $\rho_i$ is the ion mass density. 
According to the anisotropic scaling of Alfv\'{e}nic turbulence (Eq. \eqref{eq: turani}), 
the perpendicular neutral-ion decoupling scale is, 
\begin{equation}\label{eq: genkdec}
  l_{\text{ni, dec}, \perp} = \nu_\mathrm{ni}^{-\frac{3}{2}} L_\text{st}^{-\frac{1}{2}} V_\text{st}^{\frac{3}{2}}. 
\end{equation}
Since the ion-neutral collisional damping of Alfv\'{e}nic turbulence is weak when the neutral-ion coupling is strong 
\citep{XuL15,Xuc16,XLr17}, 
here we consider that the damping effect is insignificant on length scales down to $l_{\text{ni, dec}, \perp}$ and 
thus apply the anisotropic scaling of quasi-single-fluid Alfv\'{e}nic turbulence to determine $l_{\text{ni, dec}, \perp}$.
On scales larger than $l_{\text{ni, dec}, \perp}$,
the strongly coupled neutrals and ions together carry the 
anisotropic MHD turbulence, and density fluctuations present a filamentary structure along the magnetic field due to the 
perpendicular turbulent mixing as discussed above. 
On smaller scales, 
neutrals decouple from ions and the magnetic field. 
Therefore they independently carry the isotropic hydrodynamic turbulence
\citep{XuL15, Burk15}. 
In the case when the viscous damping scale of turbulence is sufficiently small,
the cascade of MHD turbulence in neutrals 
terminates at $l_{\text{ni, dec}, \perp}$, which is thus the minimum width of low-density 
parallel filaments in the neutral gas.
We note that due to the turbulent energy cascade and turbulent mixing of density fluctuations, 
the density fluctuations at $l_{\text{ni, dec}, \perp}$ are relatively low, and thus 
we do not expect that $l_{\text{ni, dec}, \perp}$ characterizes a typical scale of density structures.

We should stress that the ``ambipolar diffusion (AD) length" frequently adopted in the literature 
(e.g., \citealt{Mous91,Bals96,Nto16})
is in fact $l_{\text{ni, dec}, \|}$ in Eq. \eqref{eq: panid}. 
Its formulation is based on a linear description of Alfv\'{e}n waves. 
For the nonlinear anisotropic Alfv\'{e}nic turbulence, only $l_{\text{ni, dec}, \perp}$ in 
Eq. \eqref{eq: genkdec} is relevant 
\citep{XuL15,Xuc16}.
The disparity between $l_{\text{ni, dec}, \|}$ and $l_{\text{ni, dec}, \perp}$ can be quite large especially in sub-Alfv\'{e}nic turbulence.
The latter is difficult to be resolved numerically due to the limited numerical resolution. 
The two-fluid MHD simulations by 
\citet{Burk15}
showed evidence for the existence of Alfv\'{e}nic turbulence on scales below $l_{\text{ni, dec}, \|}$. 
More high-resolution two-fluid numerical tests on the behavior of MHD turbulence at $l_{\text{ni, dec}, \|}$ and $l_{\text{ni, dec}, \perp}$
are necessary.

To illustrate the filamentary structure in the partially ionized ISM, 
we adopt the typical driving conditions of the interstellar 
turbulence, $L = 30$ pc, $V_L = 10$ km s$^{-1}$, 
and typical parameters of the WNM, 
the CNM, and MCs
\citep{Dr98} (see Table \ref{tab: cnm}). 
Besides, we use $\gamma_d=5.5\times10^{14}$cm$^3$g$^{-1}$s$^{-1}$,
the ion and neutral masses $m_\mathrm{i} = m_\mathrm{n} = m_\mathrm{H}$ for the WNM and the CNM, 
and $\gamma_d=3.5\times10^{13}$cm$^3$g$^{-1}$s$^{-1}$, $m_\mathrm{i} = 29 m_\mathrm{H}$, $m_\mathrm{n} = 2.3 m_\mathrm{H}$ for MCs 
\citep{Drai83,Shu92}, 
where $m_\mathrm{H}$ is the mass of hydrogen atom.

In the case of the WNM, given the above parameters, the turbulence is sub-Alfv\'{e}nic. 
The viscous damping scale
\begin{equation}\label{eq: nuds}
   l_{\nu,\perp} = \nu_n^\frac{3}{4} L^\frac{1}{4} V_L^{-\frac{3}{4}} M_A^{-\frac{1}{4}},
\end{equation}
is larger than $l_{\text{ni, dec}, \perp}$, 
where $\nu_n = v_\text{th}/ (n_n \sigma_{nn})$ is the viscosity in neutrals, 
with the neutral thermal speed $v_\text{th}$, the neutral number density $n_n$, 
and the cross-section of a neutral-neutral collision $\sigma_{nn} \sim 10^{-14} ~\text{cm}^2$
\citep{VrKr13}. 
It means that the sub-Alfv\'{e}nic turbulence in the WNM is damped due to neutral viscosity  
when ions and neutrals are strongly coupled
\citep{LVC04,Xuc16}.
So the density structure of both neutral and ionized gases at $l_{\nu,\perp}$ exhibit the anisotropy of MHD turbulence. 
To quantify the elongation of the viscous-scale density filament, we calculate the axial ratio as (Eqs. \eqref{eq: subalf}, \eqref{eq: geraas} and \eqref{eq: nuds}, Table \ref{tab: cnm})
\begin{equation}
    A_{\nu,\text{sub}} = \frac{l_{\nu,\|}}{l_{\nu,\perp}} = \nu_n^{-\frac{1}{4}} L^\frac{1}{4} V_L^\frac{1}{4} M_A^{-\frac{5}{4}},
\end{equation}
which reflects the local turbulence anisotropy and
has a large value in the sub-Alfv\'{e}nic WNM.

In cases of the CNM and MCs, with the above parameters adopted, the turbulence turns out to be super-Alfv\'{e}nic.
Different from the WNM, 
there is $l_{\text{ni, dec}, \perp} >  l_{\nu,\perp}$ in the CNM and MCs, 
so the effect of neutral-ion decoupling on MHD turbulence is important.
To determine the neutral-ion decoupling scale, 
we can rewrite Eq. \eqref{eq: genkdec} by using Eq. \eqref{eq: supalf},
\begin{equation}\label{eq: desups}
  l_{\text{ni, dec}, \perp, \text{sup}} = \nu_\mathrm{ni}^{-\frac{3}{2}} L^{-\frac{1}{2}} V_L^{\frac{3}{2}}  ,
\end{equation}
which is independent of the magnetic field strength.
The axial ratio at the decoupling scale is 
(Eqs. \eqref{eq: panid} and \eqref{eq: desups})
\begin{equation}
     A_\text{dec, sup} = \frac{l_{\text{ni, dec}, \|}}{l_{\text{ni, dec}, \perp, \text{sup}}} = \nu_\mathrm{ni}^\frac{1}{2} L^\frac{1}{2} V_L^{-\frac{1}{2}} \mathcal{M}_\mathrm{A}^{-1}.
\end{equation}
The reason for the small $A_\text{dec, sup}$ in a highly super-Alfv\'{e}nic MC is that 
with comparable turbulent and magnetic energies at $L_\text{st}$ and with $l_{\text{ni, dec}, \perp, \text{sup}}$ close to $L_\text{st}$, 
the turbulence anisotropy at $l_{\text{ni, dec}, \perp, \text{sup}}$ is insignificant. 

Obviously, 
compared with the super-Alfv\'{e}nic CNM and MCs, 
we see that the sub-Alfv\'{e}nic WNM is more favorable for the formation of 
profoundly elongated low-density parallel filaments.

\begin{table*}[t]
\centering
\begin{threeparttable}
\caption[]{Parameters in different partially ionized interstellar phases}\label{tab: cnm} 
  \begin{tabular}{cccccccccccc}
     \toprule
              &  $n_H [\text{cm}^{-3}]$ &  $n_e/n_H$         & $B_0$ [$\mu$ G] &  $T$ [K]  & $\mathcal{M}_\mathrm{s}$    & $\mathcal{M}_\mathrm{A}$    &  $L_\text{st}$ [pc]    &  $l_{\nu,\perp}$ [pc] & $l_{\text{ni, dec}, \perp}$ [pc]  &  $A_\nu$  & $A_\text{dec}$  \\
      \hline
   WNM  & $0.4$                          &  $0.1$                & $5$           &    $6000$  &  $1.1$         &  $0.6$      &   $11.1$                         &  $2.3\times10^{-3}$    & -   &    $46.0$  & -      \\
   CNM  &  $30$                           &  $10^{-3}$         &  $5$           &    $100$    &  $8.5$         &     $5$       &   $0.2$                          & -  &    $2.3 \times10^{-4}$   &-       &  $10.0$    \\
   MC     &  $300$                         &  $10^{-4}$         & $5$            &    $20$     &   $28.9$       &     $15.9$   &   $7.5\times10^{-3}$    &- &  $9.3\times10^{-5}$      &-        &  $4.3$   \\
    \bottomrule
    \end{tabular}
 \end{threeparttable}
\end{table*}

\subsection{Examples for low-density parallel filaments in the ISM}
\label{ssec: obspaf}

Perpendicular turbulent mixing of Alfv\'{e}n modes 
can account for the formation of low-density parallel filaments in both diffuse warm phases with subsonic to transonic turbulence 
and dense cold phases with supersonic turbulence. Some examples are as follows.

(1) {\it HI filaments.} Observations show that filamentary structures in neutral hydrogen (HI) gas with a 
median column density $N_\mathrm{HI} \approx 10^{19.1}$ cm$^{-2}$ are aligned with the magnetic field measured by Planck
(\citealt{Kal16}; see also \citealt{Cl14,Cl15}). 
Here we caution that to extract density structures from channel maps, the velocity dispersion caused by turbulence should be taken into account and 
thus a sufficient channel spacing should be used 
\citep{Laz09rev}.
Direction-dependent HI power spectra associated with HI filaments show turbulence anisotropy 
\citep{Kalb16,Kal17},
which we suggest as the origin of low-density parallel filaments.
We note that the GS95 turbulence anisotropy is only observable in the reference frame of the {\it local} magnetic field 
(LV99; \citealt{CLV_incomp}), 
which is different from the anisotropy obtained in the global reference frame involving the projection effect.

(2) {\it Striations in diffuse regions of MCs.}  
Striations, a network of faint and elongated density structures aligned along the magnetic field, are found 
in the low-column density ($N_H \approx 1-2 \times10^{21}$ cm$^{-2}$) regions in MCs
\citep{Gol08,Hey16}.
The high velocity gradient transverse to the magnetic field compared to that along the field direction 
\citep{Hey08}
and the spatial power spectrum transverse to the magnetic field 
\citep{Tri16}
provide strong observational evidence for anisotropic turbulent mixing as the mechanism of their formation. 

(3) {\it Subfilaments in MCs.} Subfilaments are the striations perpendicular and connected to the main filaments,
which can contribute to the growth of main filaments through the mass accretion along them
\citep{Gol08,Sch10,Hen12, Pal13}.
In this scenario, the velocity distribution within subfilaments is mainly 
regulated by gravitational infall instead of turbulent mixing.

\section{Formation of dense filaments in supersonic interstellar turbulence }
\label{sec: densho}

In cold MCs, the turbulence is highly supersonic 
\citep{Zuck74,Lars81,HB04}.
Apart from the low-density parallel filaments induced by the perpendicular mixing of Alfv\'{e}nic turbulence (Section \ref{sec:fldpa}), 
a network of shocks driven by supersonic turbulent flows generate 
significant density enhancements
\citep{Pad04,MacL04,KL07}. 
The shock compression deforms the spectrum of density fluctuations to be different from that of turbulent velocities
\citep{BLC05,KL07}
and accounts for the formation of dense filaments in MCs
\citep{Rob18,Moc18}\footnote{In 3D supersonic turbulence, 
due to the interaction of density inhomogeneities with multiple shocks and 
turbulent mixing 
\citep{Pan10},
filamentary structures instead of 2D sheet-like structures are more likely to be generated. }.


The density contrast across the shock is 
\begin{equation}\label{eq: shm}
    \frac{\rho_2}{\rho_1} \approx \sqrt{2} \mathcal{M}_\mathrm{A1}
\end{equation}
when the magnetic pressure dominates in the downstream region, 
and 
\begin{equation}\label{eq: sht}
    \frac{\rho_2}{\rho_1} \approx  \mathcal{M}_\mathrm{s1}^2
\end{equation}
when the thermal pressure dominates in the downstream region, 
where $\rho_1$ and $\rho_2$ are densities of the upstream and downstream media, 
$\mathcal{M}_\mathrm{A1}$ is the shock Alfv\'{e}n Mach number corresponding to the 
transverse component of upstream magnetic field, and $\mathcal{M}_\mathrm{s1}$ is the shock sonic Mach number
(see Appendix \ref{app: de} for the detailed derivation).
The above compression ratios 
are consistent with earlier studies on isothermal shocks 
(see, e.g., \citealt{Draine:2011aa}). 
In highly supersonic MHD turbulence, the shock-compressed density varies in different turbulence regimes.

(1) {\it $\mathcal{M}_\mathrm{A} \gg \mathcal{M}_\mathrm{s} \gg 1$.}  
In a weakly magnetized medium, the shock compression in MHD turbulence is similar to that in 
hydrodynamic turbulence. 
The resulting density enhancement solely depends on $\mathcal{M}_\mathrm{s1}(\sim \mathcal{M}_\mathrm{s})$
(Eq. \eqref{eq: sht}, see \citealt{Pad11,Fed15,Fed16,Rob18}).
We note that the weak magnetic field can be amplified by the downstream turbulence 
\citep{XL16,XL17}.

(2) {\it $\mathcal{M}_\mathrm{s} \gg \mathcal{M}_\mathrm{A} > 1$} or {\it $\mathcal{M}_\mathrm{s} \gg 1 > \mathcal{M}_\mathrm{A}$.}
In a strongly magnetized medium, 
Eq. \eqref{eq: shm} applies to the shock compression of an oblique shock. 
Only for a quasi-parallel shock, Eq. \eqref{eq: sht} can be used for describing the shock-compressed density. 

We can clearly see the magnetic effect on the formation of dense filaments. 
In the weak-field limit, with no preferred shock propagation direction for the density enhancement, 
the generated filaments with similar densities are randomly orientated. 
In the strong-field limit, 
quasi-parallel shocks are more favorable for generating a large density contrast, 
resulting in dense filaments nearly perpendicular to magnetic fields.
The numerical evidence can be found in, e.g., 
\citet{KL07,Moc18}.
The preference on dense perpendicular filaments in observations of MCs 
\citep{Pla16,Pln16}
indicates the dominance of magnetic energy over thermal energy. 
With respect to the turbulent energy, 
we note that even the turbulence is super-Alfv\'{e}nic at the large driving scale, 
the magnetic energy can still exceed the turbulent energy on smaller scales 
due to the turbulent energy cascade.

In addition, the width of dense filaments can be estimated as the shock propagation distance reduced by the 
compression ratio. 
For an oblique filament, the width can be estimated as (Eq. \eqref{eq: shm}),
\begin{equation}
     \delta_{o} \sim  \frac{L}{\sqrt{2} \mathcal{M}_\mathrm{A1}},
\end{equation}
which depends on the magnetization and the filament orientation.
For the densest perpendicular filaments, 
by assuming a constant compression ratio determined by 
$\mathcal{M}_\mathrm{s1} (\sim \mathcal{M}_\mathrm{s})$, we can approximately have the filament width as (Eq. \eqref{eq: sht})
\begin{equation}\label{eq:shtn}
   \delta_\perp \sim \frac{L}{\mathcal{M}_\mathrm{s}^2},
\end{equation} 
which depends on the 
compressibility of medium. 
Provided $L \sim 10$ pc and $\mathcal{M}_\mathrm{s} \sim 10$
\citep{Sch13,Hen16}, 
we find that $\delta_\perp$ is of the order of $0.1$ pc, consistent with observations by,
e.g., \citet{Arz11,Koc15}.

\section{Discussion}


Motivated by ample observations of interstellar density filaments, 
there have been numerous theoretical arguments and numerical experiments on their formation mechanisms 
(e.g., \citealt{Laz93,Henn13,Smi14,Nto14,Ino16,Smi16,Fed16,Ban18}).
In this work, based on the fundamental dynamics of Alfv\'{e}nic turbulence, 
we propose a formation mechanism of low-density parallel filaments in the ISM.

{\it Cold HI filaments in the WNM.}
Based on numerical studies, 
\citet{Sau14} 
argued that the dynamics of the WNM and the CNM are tightly coupled, 
and the velocity dispersion between CNM clumps is close to the velocity dispersion within the extended WNM. 
It implies that the denser CNM embedded in the diffuse WNM is dynamically mixed by the turbulent velocities in the WNM, 
and thus the cold HI filaments are aligned with the turbulent magnetic fields in the WNM
\citep{Kal16}.
In this scenario, the elongation of cold HI filaments is determined by the turbulence anisotropy in the WNM
(see Section \ref{ssec: niani}).
The single-phase MHD simulations performed in this work are inadequate to capture the distinctive turbulence properties of the WNM and CNM 
and their interactions. 
As we have mentioned, 
to fully understand the turbulence dynamics in the multi-phase ISM, a model for (at least) two-phase MHD turbulence should be applied.

{\it Gravitational effects on dense filaments in MCs.}
Dense filaments can form in supersonic turbulence without gravity, instead they are 
the seeds of gravitationally collapsing regions
(e.g., \citealt{Fed13,Rob18,Moc18}).
Dense filaments in MCs set up the necessary condition for the self-gravity to take over the 
gas dynamics and initiate the subsequent star formation. 
Gravitational effects, which were not considered in this work, 
can redistribute the spatial structure of striations, 
narrow the filament width, 
and change the velocity distribution and magnetic field orientation within filaments
\citep{Chi17}.


{\it Relation of density filaments to the density gradient (DG) and velocity gradient (VG).}
For a filamentary density structure, the DG is perpendicular to its major axis.
Accordingly, the DG of a parallel filament is perpendicular to the magnetic field ${\bf B}$, 
while that of a perpendicular filament is parallel to ${\bf B}$. 
In a good agreement with our analysis, 
\citet{Yu17}
found that DGs measured in HI gas are perpendicular to ${\bf B}$, 
and 
\citet{YuL17}
showed DGs parallel to ${\bf B}$ at shocks. 
But we caution that the DG is not always a good indicator of density filaments. 
For instance, DGs parallel to ${\bf B}$ can also arise in a gravitationally collapsing region. 
On the other hand, due to the perpendicular turbulent mixing (Section \ref{ssec: ptmpf}), 
the VG of Alfv\'{e}nic turbulence is perpendicular to ${\bf B}$
\citep{Gon17}.
The low-density parallel filaments and the perpendicular VGs observed in both HI gas and MCs 
\citep{Yu17,LY18}
have the same physical origin.

It is also important to note that 
in channel maps with a small channel spacing, the intensity fluctuations are mostly influenced by velocity fluctuations. 
Density filaments can only be extracted from sufficiently thick velocity channel maps 
\citep{LP00,LY18}.

\section{Summary}

The pervasive magnetized turbulence is responsible for the formation of ubiquitous density filaments in the ISM. 
Corresponding to different turbulence regimes in different interstellar phases, the resulting filaments have distinctive features. 
Our main results are summarized as follows.

Based on our theoretical understanding of MHD turbulence and numerical experiments, we showed that in 
compressible MHD turbulence,
the perpendicular mixing by Alfv\'{e}nic turbulence gives rise to 
low-density filaments aligned with the local magnetic field, 
which explains the low-density parallel filaments observed in diffuse interstellar media and low-density regions of MCs. 
The filament elongation 
depends on the anisotropy of Alfv\'{e}nic turbulence and varies in different turbulence regimes. 
More elongated low-density parallel filaments are expected in magnetic energy-dominated turbulence. 
In the CNM and MCs, 
the minimum width of a low-density parallel filament is determined by the perpendicular 
neutral-ion decoupling scale.

In the highly supersonic MHD turbulence in MCs, shock compression generates dense filaments. 
The filament width depends on the relative importance between magnetic and thermal energies, 
as well as shock propagation direction. 
\\
\\

S.X. acknowledges the support for Program number HST-HF2-51400.001-A provided by NASA through a grant from the Space Telescope Science Institute, which is operated by the Association of Universities for Research in Astronomy, Incorporated, under NASA contract NAS5-26555.
S.X. also thanks Chris McKee for useful conversations.
S.J. acknowledges the support from the Sherman Fairchild Fellowship at Caltech.
A.L. acknowledges the support from grant NSF DMS 1622353.
Simulations are preformed on Blue Waters supercomputer at NCSA, under the allocation PRAC NSF.1713353 supported by NSF. We have made use of NASA's Astrophysics Data System and the yt astrophysics analysis software suite \citep{turk2010yt}.

\appendix
\section{Shock compression in highly supersonic MHD turbulence}\label{app: de}

In the rest frame of the shock, the quantities across the shock are related by the jump conditions, which include the 
conservation of mass, 
\begin{equation}\label{eq: conm}
    \rho_1 u_1  = \rho_2 u_2,
\end{equation}
the conservation of momentum, 
\begin{equation}\label{eq: csmo}
    \rho_1 c_1^2 + \rho_1 u_1^2 + \frac{B_1^2}{8\pi} =  \rho_2 c_2^2 + \rho_2 u_2^2 + \frac{B_2^2}{8\pi},
\end{equation}
and the conservation of magnetic flux 
\begin{equation}\label{eq: csfl}
       B_1 u_1 =  B_2 u_2, 
\end{equation}
where $\rho_1$, $u_1$, $c_1$, $B_1$ are the mass density, fluid velocity (in the shock propagation direction),
the sound speed, the strength of the transverse magnetic field in the upstream region, 
and $\rho_2$, $u_2$, $c_2$, $B_2$ are those in the downstream region.

Under the consideration of highly supersonic turbulence, Eq. \eqref{eq: csmo} is approximately 
\begin{equation}
   \rho_1 u_1^2 + \frac{B_1^2}{8\pi}  \approx  \rho_2 c_\mathrm{s}^2 + \rho_2 u_2^2 + \frac{B_2^2}{8\pi}.
\end{equation}
Here we also assume that the shock is isothermal with $c_1 = c_2 = c_\mathrm{s}$ due to the efficient cooling in MCs 
(see e.g. \citealt{Wh97}).
In combination with Eqs. \eqref{eq: conm} and \eqref{eq: csfl}, the above equation becomes
\begin{equation}
      \Bigg(1-\frac{V_\mathrm{A2}^2}{2u_1^2} \Bigg) u_2^2  - u_1 u_2 +  c_\mathrm{s}^2 + \frac{V_\mathrm{A2}^2}{2}  \approx 0,
\end{equation}
as a quadratic equation of $u_2$, 
where $V_\mathrm{A2} = B_2 /\sqrt{4\pi\rho_2}$. 
Its solutions are: 
\begin{equation}\label{eq:sol}
     u_2 \approx \frac{u_1 \pm u_1\sqrt{1 - 4 \Big(1-\frac{V_\mathrm{A2}^2}{2 u_1^2}\Big) \Big(\frac{c_\mathrm{s}^2}{u_1^2} + \frac{V_\mathrm{A2}^2}{2 u_1^2} \Big)}}{2 - \frac{V_\mathrm{A2}^2}{u_1^2}}  .
\end{equation}
Based on this general form, we next discuss the shock compression in the following two cases:  
 
{\it Case~1:} Magnetic-pressure dominated downstream medium

At $ c_\mathrm{s}^2 \ll  V_\mathrm{A2}^2 $,
Eq. \eqref{eq:sol} can be simplified as 
\begin{equation}
    u_2 \approx \frac{u_1 \pm u_1 \Big(1-\frac{V_\mathrm{A2}^2}{u_1^2}\Big)}{2},
\end{equation}
where $V_\mathrm{A2}^2/u_1^2 \ll 1$ should be satisfied. 
We consider the solution corresponding to non-negligible compression of the shocked material and obtain, 
\begin{equation}\label{eq: uudrd}
    u_2 \approx \frac{V_\mathrm{A2}^2}{2u_1}.
\end{equation}
From Eqs. \eqref{eq: conm} and \eqref{eq: csfl}, we find $V_\mathrm{A2}^2 = (\rho_2/\rho_1) V_\mathrm{A1}^2$, where $V_\mathrm{A1} = B_1 /\sqrt{4\pi\rho_1}$,
and that the relation in Eq. \eqref{eq: uudrd} determines the density contrast across the shock as
\begin{equation}\label{eq:crsal}
    \frac{\rho_2}{\rho_1} \approx \sqrt{2} \frac{u_1}{V_\mathrm{A1}} = \sqrt{2} \mathcal{M}_\mathrm{A1},
\end{equation}
where $\mathcal{M}_\mathrm{A1}$ is the shock Alfv\'{e}n Mach number corresponding to the 
transverse component of upstream magnetic field.

{\it Case~2:} Thermal-pressure dominated downstream medium

At $c_\mathrm{s}^2 \gg  V_\mathrm{A2}^2$, the solutions in Eq. \eqref{eq:sol} can be reduced to 
\begin{equation}
      u_2 \approx \frac{u_1 \pm u_1 \sqrt{1- \frac{4 c_\mathrm{s}^2}{u_1^2}}}{2}. 
\end{equation}
We again only consider the situation with a significant compression and find
\begin{equation}
     u_2 \approx \frac{c_\mathrm{s}^2}{u_1}
\end{equation}
for a supersonic $u_1$. 
Combining the above expression with Eq. \eqref{eq: conm} yields
\begin{equation}\label{eq:crsm}
    \frac{\rho_2}{\rho_1} \approx \frac{u_1^2}{c_\mathrm{s}^2}  = \mathcal{M}_\mathrm{s1}^2,
\end{equation}
where $\mathcal{M}_\mathrm{s1}$ is the shock sonic Mach number.

\bibliographystyle{apj.bst}
\bibliography{xu}

\end{document}